\documentclass[useAMS,usenatbib,twocolumn]{mnras}
\usepackage{epsfig,amssymb}
\usepackage{graphicx,amsmath,multicol}
\usepackage{multirow}
\usepackage{caption}
\usepackage{subcaption}
\usepackage{placeins}
\usepackage{physics}

\usepackage[utf8]{inputenc}
\usepackage{setspace}
\usepackage{graphicx}
\usepackage{mathtools}
\usepackage{grffile}
\usepackage{amsmath}
\usepackage{longtable}
\usepackage{graphics}
\usepackage{verbatim}
\usepackage[usenames]{color}
\usepackage{float}
\title[GRB prompt emission radiation mechanism]{Explaining GRB prompt emission with sub-photospheric dissipation and Comptonization}
\author[Bhattacharya and Kumar]
{Mukul Bhattacharya$^{1}$\thanks{E-mail: mukul.b@utexas.edu (MB)} 
and Pawan Kumar$^2$\thanks{pk@astro.as.utexas.edu}\\  
$^{1}$ Department of Physics, University of Texas at Austin, Austin, TX 78712, USA\\
$^{2}$ Department of Astronomy, University of Texas at Austin, Austin, TX 78712, USA}
\begin{document}

\date{Accepted . Received ; in original form }

\pagerange{\pageref{firstpage}--\pageref{lastpage}} \pubyear{2019}

\maketitle

\label{firstpage}

\begin{abstract}
Even though the observed spectra for GRB prompt emission is well constrained, no single radiation mechanism can robustly explain its distinct non-thermal nature. Here we explore the radiation mechanism with the photospheric emission model using our Monte Carlo Radiative Transfer (MCRaT) code. We study the sub-photospheric Comptonization of fast cooled synchrotron photons while the Maxwellian electrons and mono-energetic protons are accelerated to relativistic energies by repeated dissipation events. Unlike previous simulations, we implement a realistic photon to electron number ratio $N_{\gamma}/N_{e} \sim 10^5$ consistent with the observed radiative efficiency of a few percent. We show that it is necessary to have a critical number of episodic energy injection events $N_{rh,cr} \sim {\rm few}\ 10{\rm s}-100$ in the jet in addition to the electron-proton Coulomb coupling in order to inject sufficient energy $E_{inj,cr} \sim 2500-4000\ m_e c^2$ per electron and produce an output photon spectrum consistent with observations. The observed GRB spectrum can be generated when the electrons are repeatedly accelerated to highly relativistic energies $\gamma_{e,in} \sim {\rm few}\ 10{\rm s}-100$ in a jet with bulk Lorentz factor $\Gamma \sim 30-100$, starting out from moderate optical depths $\tau_{in} \sim 20-40$. The shape of the photon spectrum is independent of the initial photon energy distribution and baryonic energy content of the jet and hence independent of the emission mechanism, as expected for photospheric emission.
\end{abstract}


\section{Introduction}
\label{Intro}
%\vspace{0.05in}

The radiation mechanism responsible for long-duration Gamma-Ray Burst (GRB) prompt emission has remained elusive ever since their discovery five decades ago. The observed spectrum has a distinctly non-thermal shape and is often modelled using the Band function with a smoothly connected broken power-law shape (\citealt{Band}). While the observed peak photon energy is at $E_{\rm{peak}} \sim 300$ keV, the low/high energy spectrum is given by the power-law $f_{\nu} \propto \nu^{0}$/$f_{\nu} \propto \nu^{-1.2}$ \citep{Preece,Kaneko,Kaneko08}. A robust radiation mechanism should explain all these features of the prompt emission spectrum in a self-consistent manner. Synchrotron and photospheric models are the two most widely studied models to this end (see \citealt{Piran04,KZ15} for detailed reviews).

In the synchrotron model, electrons accelerated to relativistic energies either by internal shocks \citep{RM94} or magnetic reconnection \citep{Gia06} produce the prompt radiation via synchrotron emission process \citep{Meszaros94,Piran99}. While this model accounts for the broad non-thermal nature of the prompt spectrum, it cannot explain the high radiation efficiencies confirmed by observations \citep{Zhang07}. Another shortcoming of this model is that the observed low-energy hard spectrum cannot be accounted for by the synchrotron emission process \citep{Preece98,Ghirlanda03}.
However, \citet{Uhm14} and \citet{Geng18} have recently shown that the hardening of the low-energy GRB prompt emission spectrum can possibly be explained with a gradually decreasing magnetic field strength in the emission region.

These difficulties with the synchrotron model have led researchers to consider photospheric emission model in more detail \citep{MR00,RM05,CL15,MB18}. The photospheric model naturally explains the high radiation efficiencies for prompt emission without assuming any specific dissipation mechanism. Furthermore, the observed spectrum is completely determined by the electron-photon interaction in the jet irrespective of the dissipation mechanism involved. While the high-energy non-thermal behaviour has been successfully explained by sub-photospheric dissipation \citep{Gia06,LB10,Vurm11,CL15,MB18}, reproducing the low-energy non-thermal tails has turned out to be really challenging \citep{LB10,CL15,MB18}.

In this paper, we study the photospheric emission model in further detail to find the plausible conditions under which both the low/high-energy non-thermal behaviour and the observed peak energy can be explained self-consistently. We consider a radiation-matter coupling via Comptonization i.e. photons undergoing multiple scatterings with electrons accelerated below the photosphere. Repeated dissipation events such as internal shocks \citep{RM94,LB10,Toma11} or magnetic reconnection \citep{Thompson94,Gia06} accelerate the electrons and protons in the jet to relativistic energies as the outflow expands outwards. These highly energetic electrons then cool rapidly to generate photons with a fast-cooled synchrotron spectrum with a characteristic broken power-law shape \citep{Ghisellini00,Granot00,Piran04}. 
In this work, we define \emph{sub-photospheric} events as the physical processes such as episodic dissipation, Coulomb collisions and Comptonization which occur below the photospheric radius of the relativistic outflow and at moderate optical depths $\tau \sim {\rm few}-10{\rm s}$. 
Unlike previous studies, photons in our work initially do not have a thermal distribution as they do not undergo sufficiently many scatterings after being produced at relatively moderate $\tau \lesssim 50$ \citep{Begue13}. 

The bulk of the jet energy is contained as the kinetic energy of the protons with the average energy of the photons being much smaller compared to that of the electrons, thereby enabling significant energy transfer to the photons. In addition to the sub-photospheric episodic dissipation events, electrons are also accelerated continuously by the protons via Coulomb collisions \citep{MB18}. While the outflow is optically thick, the photons continue to scatter electrons and gain energy until either the average photon energy matches that of the electrons or the outflow becomes optically thin so that the photons escape the photosphere. The photon spectrum can get significantly broadened due to both Comptonization with energetic electrons and geometrical effects \citep{Begue13,Lundman13,MB18}. The shape of the photon spectrum changes considerably with the photon to electron number ratio $N_{\gamma}/N_{e}$ as well \citep{MB18}. For typical values of jet bulk Lorentz factor $\Gamma$ and photon peak energy $E_{\rm{peak}}$, $N_{\gamma}/N_{e} \sim 10^{5}$ for radiation efficiency $\eta \sim 10\%$ as confirmed by observations.

In this work, we present results of MCRaT simulations performed with realistic $N_{\gamma}/N_e$ values (see, \citealt{MB18}, for details on the code implementation). The initial distributions for the electrons, protons and photons are taken to be Maxwellian, mono-energetic and broken power-law, respectively. We determine a correlation between the number of reheating events $N_{\rm{rh}}$ and the initial optical depth $\tau_{\rm{in}}$ and perform an exhaustive parameter space search in order to obtain a Band-like observed spectrum. We also perform analytical calculations to examine the evolution of photon energy spectrum with multiple scatterings and validate the MCRaT simulation results.
 
This paper is organized as follows. In Section \ref{ph_e_req}, we estimate the electron energy required to produce a photon spectrum with peak energy consistent with observations and argue that continuous electron heating via electron-proton Coulomb collisions is insufficient for maintaining electrons at this energy. In Section \ref{e_rh}, we evaluate the electron energy by including the effect of adiabatic energy loss and show that the number of sub-photospheric dissipation events needed to keep electrons sufficiently hot is closely related to the optical depth where the particles and photons are injected into the jet to start interacting. In Section \ref{code_desc}, we describe the basic implementation of our photospheric MCRaT code in addition to briefly discussing the relevant physics involved. We present the MCRaT simulation results in Section \ref{sim_results} and explore the parametric space in detail to constrain the GRB prompt emission parameters. In Section \ref{IC_sc}, we analytically compute the scattered photon spectrum by assuming Comptonization as the dominant process and further show that the output photon spectrum becomes increasingly non-thermal over repeated scatterings to resemble the observed spectrum. Finally, we discuss the interpretation of the simulation results and present our conclusions in Section \ref{disc_conc}. Throughout this paper, we use primed/unprimed coordinates for jet-comoving/lab frame quantities.

%\vspace{0.1 in}
\section{Photon energy requirement}
\label{ph_e_req}
In this section, we first estimate the average energy $E_{\gamma,avg}$ that the photons in the outflow need to have in order to produce a Band-like output spectrum. Since, most of this energy is transferred by the hot electrons via Comptonization, the electrons need to have certain threshold energy $\gamma_{e,crit}$ that we then compute. The electrons can be maintained at this critical energy either by Coulomb collisions with protons or repeated dissipation events that occur while the outflow expands. We argue here that the electron-proton Coulomb coupling alone is not sufficient for supplying the bulk of the energy to the photons and sub-photospheric dissipation events are necessary to obtain Band-like GRB prompt spectrum.

\subsection{Analytical estimate for $E_{\gamma,avg}$ and $\gamma_{e,crit}$}
The observed photon spectrum has a Band-like shape with a low/high-energy dependence, $f_{\nu} \propto \nu^0/\nu^{-1.2}$ in the energy range $\sim 10\ {\rm keV} - 300\ {\rm keV}$/$\sim 300\ {\rm keV} - 10\ {\rm MeV}$ \citep{Preece,Kaneko}, where $f_{\nu}$ denotes the photon flux per unit frequency.
The average observed energy of each photon in the lab frame is then
\begin{equation}
E_{\gamma,{\rm avg}}^{{\rm obs}} = \left.\int_{10 {\rm\ keV}}^{300 {\rm\ keV}}f_{\nu}\ d\nu\middle/\int_{10{\rm\ keV}}^{300{\rm\ keV}}(f_{\nu}/\nu)\ d\nu \sim 100 {\rm\ keV}.\right.
\label{avgEph}
\end{equation}
In the jet-comoving frame (for jet bulk Lorentz factor $\Gamma = 300$), $E_{\gamma,avg}^{\prime} = 0.33$ keV.
We can now estimate how energetic the electrons have to be in order to deposit sufficient energy $\sim N_{\gamma}E_{\gamma,avg}^{obs}$ into photons after multiple scatterings.

Assuming that the electron-positron pair processes can be ignored and with $N_{\gamma} = 2\times10^{7}$, $N_{e} = N_{p} = 2\times10^{2}$ throughout, in the jet-comoving frame\\
{\it Total energy content of photons = energy that the energetic electrons deposit into the photons via Comptonization}
\begin{equation}
N_{\gamma}E_{\gamma,{\rm avg}}^{\prime} = N_{e}(\gamma_{e,{\rm crit}} - 1)m_{e}c^{2} \left(\tau_{{\rm in}}t_{{\rm dyn}}^{\prime}/t_{{\rm IC}}^{\prime}\right), \\
\label{ge_no_shock_with_supCoul}
\end{equation}
where $\tau_{in}$ is the initial optical depth, $t_{{\rm dyn/IC}}^{\prime}$ is the dynamical/inverse-Compton (IC) timescale and $\tau_{{\rm in}}t_{{\rm dyn}}^{\prime}/t_{{\rm IC}}^{\prime} \sim$ number of times the electrons interact with photons during jet expansion. Here we assume that $\gamma_{e,crit}$ and $t_{IC}^{\prime}$ 
%stay approximately constant 
do not vary significantly during jet expansion, that is the electrons remain in (approximate) equilibrium. The characteristic dynamical and IC timescales are given by
\begin{eqnarray}
t_{{\rm dyn}}^{\prime} = \frac{R_{in}}{\Gamma c} = \frac{L\sigma_T}{8\pi m_p c^4 \beta \Gamma^4 \tau_{in}},\\
t_{{\rm IC}}^{\prime} = \frac{3(\gamma_{e}-1)m_e c}{4 U_{\gamma}^{\prime}\sigma_T \gamma_e^2 \beta_e^2} = \frac{3\pi (\gamma_e -1)m_e c^2 R^2 \Gamma^2}{\sigma_T \gamma_e^2 \beta_e^2 L_{\gamma}},
\end{eqnarray}
where $R_{in}$ is the photon injection radius, $L$ is the isotropic equivalent jet luminosity, $\sigma_T$ is the Thomson cross section, $\beta = \sqrt{1- \Gamma^{-2}} \sim 1$, $U_{\gamma}^{\prime}$ is the radiation energy density and $L_{\gamma}$ is the photon luminosity. Substituting typical GRB parameters: $L = 10^{52}\ {\rm erg/s}$, $\Gamma = 300$, $L_{\gamma} = 3.2\times10^{50}\ {\rm erg/s}$ and $R \sim R_{in}\tau_{in} = \frac{L\sigma_T}{8\pi m_p c^3 \Gamma^3} = 2.17\times10^{11}\ {\rm cm}$, gives $\gamma_{e,crit} = 1.352$. 
%So, the electrons need to be maintained at $\gamma_{e,crit} = 1.352$ to transfer significant energy to the photons.
 Here we have assumed that the initial energy of the photons is negligible in comparison to the observed energy and that IC is the dominant process for electron to photon energy transfer. This is a reasonable assumption considering the fact that the adiabatic cooling timescale for photons $\sim t_{dyn}^{\prime}$ is significantly larger than the IC timescale $t_{IC}$ for typical GRB parameters. 
 
We will now examine whether the electrons can be maintained at an energy $\gamma_{e,crit} = 1.352$ by Coulomb collisions with the protons. For this we introduce an efficiency factor $\eta$ for the electron-proton Coulomb interaction and also consider the situation when $\eta > 1$ due to possible plasma instability mechanisms \citep{BC88}. These mechanisms have already been discussed previously in the literature in the context of single-temperature hot accretion flows \citep{Yuan06,YN14}.
 %({\bf ref, mention briefly about accretion disk context}).

\begin{table*}
\begin{center}
\caption{\small $\gamma_{e}$ and $E_{inj}/m_e c^2 = (\gamma_{e,in} - 1)N_{rh}$ for different values of $\tau_{in}$, with adiabatic cooling}
\begin{tabular}{|| c || c ||}
\hline
\hline
\centering
$\gamma_{e} \sim \sqrt{1+1.2\ \tau_{in}^{1/3}},\ L_{\gamma} = 10^{50}\ \rm{erg/s},\ \eta(\tau_{in}) \sim 1$ & $\gamma_{e} \sim \sqrt{1+0.12\ \tau_{in}^{1/3}},\ L_{\gamma} = 10^{51}\ \rm{erg/s},\ \eta(\tau_{in}) \sim 1$ \\ \hline
\hline
$\tau_{in} = 10,\ \gamma_{e} \sim 1.893 \implies (\gamma_{e,in} - 1)N_{rh} \sim 23.175$ & $\tau_{in} = 10,\ \gamma_{e} \sim 1.12 \implies (\gamma_{e,in} - 1)N_{rh} < 0$  \\ \hline
$\tau_{in} = 20,\ \gamma_{e} \sim 2.063\implies (\gamma_{e,in} - 1)N_{rh} \sim 61.525$ & $\tau_{in} = 20,\ \gamma_{e} \sim 1.15 \implies (\gamma_{e,in} - 1)N_{rh} < 0$  \\ \hline
$\tau_{in} = 30,\ \gamma_{e} \sim 2.175 \implies (\gamma_{e,in} - 1)N_{rh} \sim 112.757$ & $\tau_{in} = 30,\ \gamma_{e} \sim 1.17 \implies (\gamma_{e,in} - 1)N_{rh} \sim 11.230$  \\ \hline
$\tau_{in} = 50,\ \gamma_{e} \sim 2.328 \implies (\gamma_{e,in} - 1)N_{rh} \sim 247.212$ & $\tau_{in} = 40,\ \gamma_{e} \sim 1.19 \implies (\gamma_{e,in} - 1)N_{rh} \sim 36.013$  \\ \hline
$\tau_{in} = 75,\ \gamma_{e} \sim 2.462 \implies (\gamma_{e,in} - 1)N_{rh} \sim 464.382$ & $\tau_{in} = 50,\ \gamma_{e} \sim 1.20 \implies (\gamma_{e,in} - 1)N_{rh} \sim 72.116$  \\ \hline
$\tau_{in} = 100,\ \gamma_{e} \sim 2.563 \implies (\gamma_{e,in} - 1)N_{rh} \sim 727.272$ & $\tau_{in} = 100,\ \gamma_{e} \sim 1.25 \implies (\gamma_{e,in} - 1)N_{rh} \sim 115.642$  \\ \hline
\hline
\end{tabular}
\end{center}
\label{withad}
\end{table*}

\subsection{How large should $\eta$ be and how fast do the protons cool?}
Here we estimate the value of the super-Coulomb efficiency parameter $\eta$ such that $\gamma_{e} \sim 1.352$. Assuming equilibrium between electron heating (Coulomb) and cooling (IC) processes over the jet expansion timescale,\\
{\it Timescale in which electrons get heated by protons ($t_{Coul}$) = timescale in which electrons get cooled by photons ($t_{IC}$)}
\begin{equation}
\frac{(\gamma_{e} - 1)m_{e}c^{2}}{5\times10^{-19}n_{e}^{\prime}}.\ \frac{(8.3\times10^{-15}T_{e}^{\prime 3/2} + \beta_{p}^{3})}{\beta_{p}^{2}}.\ \frac{1}{\eta} = \frac{3}{4}\frac{(\gamma_{e} - 1)m_{e}c}{U_{\gamma}^{\prime}\sigma_{T}\gamma_{e}^{2}\beta_{e}^{2}},
\label{eqn_super_Coul}
\end{equation}
where electron density $n_{e}^{\prime} = L/(4\pi R^2 m_{p}c^{3} \Gamma^{2}) = 4.17\times10^{15}\ {\rm cm^{-3}}$, radiation energy density $U_{\gamma}^{\prime} = L_{\gamma}/(4\pi R^{2}\Gamma^{2}c) = 2\times10^{11}\ {\rm erg/cm^{3}}$ and $T_{e}^{\prime} = \frac{1}{k_{B}}(\gamma_{e,ad} - 1)(\gamma_{e} - 1)m_{e}c^{2} = 1.98\times10^{9}\left(\gamma_{e} - 1/\gamma_{e}\right)$ is the electron temperature for a Maxwellian distribution. Here, $\gamma_{e,ad} = (4\gamma_e +1)/(3\gamma_e)$ is the adiabatic index of the electrons and $\beta_p$ is the speed of protons divided by the speed of light. We can then rewrite equation (\ref{eqn_super_Coul}) as
\begin{center}
$\eta = 2.55(\gamma_{e}^{2} - 1)\left[\frac{0.73(\gamma_{e} - 1/\gamma_{e})^{3/2} + \beta_{p}^{3}}{\beta_{p}^{2}}\right]$.
\end{center}
Substituting $\gamma_{p} \sim 1.123$ for $\tau_{in}=8$ \citep{MB18}, and $\gamma_{e} = \gamma_{e,crit}= 1.352$, we get $\eta = 4.5$.

%\subsection{Do the protons cool down too fast?}
We will now check whether super-Coulomb efficiency parameter $\eta \sim 4.5$ is physical.
The electrons cannot be continuously heated by Coulomb collisions if the protons cool down to energies comparable to that of electrons within time $t^{\prime} \sim t_{dyn}^{\prime}$. 

While the protons cool down due to Coulomb collisions and adiabatic expansion, the electrons gain energy through Coulomb and get cooled due to adiabatic cooling and IC. The electrons cannot be heated any further when,\\
{\it Total proton energy ($E_{p,tot}$) - proton energy loss due to Coulomb ($\Delta E_{p,Coul}$) - proton energy loss due to adiabatic expansion ($\Delta E_{p,ad}$) = Total electron energy ($E_{e,tot}$) + electron energy gain due to Coulomb ($\Delta E_{e,Coul}$) - electron energy loss due to adiabatic expansion ($\Delta E_{e,ad}$) - electron energy loss due to IC ($\Delta E_{e,IC}$)}, 
\begin{eqnarray}
N_{p}(\gamma_{p}-1)m_{p}c^{2} - N_{p}\int_{0}^{t^{\prime}}\frac{5\times10^{-19}n_{e}^{\prime}\beta_{p}^{2}}{[0.73(\gamma_{e} - 1/\gamma_{e})^{3/2} + \beta_{p}^{3}]}\eta dt^{\prime} \nonumber \\
- N_{p}\int_{0}^{t^{\prime}}\frac{(\gamma_{p}-1)m_{p}c^{2}}{R/\Gamma c}dt^{\prime} =  \nonumber \\
N_{e}(\gamma_{e}-1)m_{e}c^{2} + N_{e}\int_{0}^{t^{\prime}}\frac{5\times10^{-19}n_{e}^{\prime}\beta_{p}^{2}}{[0.73(\gamma_{e} - 1/\gamma_{e})^{3/2} + \beta_{p}^{3}]}\eta dt^{\prime} \nonumber \\
- N_{e}\int_{0}^{t^{\prime}}\frac{(\gamma_{e}-1)m_{e}c^{2}}{R/\Gamma c}dt^{\prime} - N_{e}\int_{0}^{t^{\prime}}(4/3)U_{\gamma}^{\prime}\sigma_{T}(\gamma_{e}^{2} - 1)cdt^{\prime}.
\end{eqnarray}
As the jet is charge neutral $N_{e} = N_{p}$ and substituting $t^{\prime} = \lambda t_{dyn}^{\prime}$ gives, 
\begin{eqnarray}
[(\gamma_{p}-1)m_{p}c^{2} - (\gamma_{e}-1)m_{e}c^{2}] - \nonumber \\
\left[\int_{0}^{\lambda t_{dyn}^{\prime}}\frac{(\gamma_{p}-1)m_{p}c^{2}}{R/\Gamma c}dt^{\prime} - 
\int_{0}^{\lambda t_{dyn}^{\prime}}\frac{(\gamma_{e}-1)m_{e}c^{2}}{R/\Gamma c}dt^{\prime}\right] \nonumber \\ 
+ \int_{0}^{\lambda t_{dyn}^{\prime}}(4/3)U_{\gamma}^{\prime}\sigma_{T}(\gamma_{e}^{2}-1)c dt^{\prime} \nonumber \\
= \int_{0}^{\lambda t_{dyn}^{\prime}}\frac{10^{-18}n_{e}^{\prime}\beta_{p}^{2}}{[0.73(\gamma_{e} - 1/\gamma_{e})^{3/2} + \beta_{p}^{3}]}\eta dt^{\prime}.
\label{ep_balance}
\end{eqnarray}
We convert the $t^{\prime}$-integral into an $R$-integral with boundary conditions: $R = R_{in}$ at $t^{\prime}=0$ and $R = R_{ph} = \tau_{in}R_{in}$ at $t^{\prime} = \tau_{in}t_{dyn}^{\prime}$, which gives $R = R_{in} + \beta c\Gamma t^{\prime}$. Here $R = R_{ph}$ is the radial distance in the lab frame at which the photons escape the photosphere. After $N_{Comp}$ scatterings, electron energy reduces to $\gamma_{e} \sim 1 + 1/(8\tau_{in})$ and proton energy $\gamma_{p} \sim \gamma_{p,in} \sim 2$ \citep{Santana16}. Substituting these values in equation (\ref{ep_balance}) and further simplification gives
\begin{center}
$(1- {\rm ln}\lambda)\left(1.5\times10^{-3} - \frac{1.02\times10^{-7}}{\tau_{in}}\right) - \int_{R_{in}}^{\lambda R_{in}}\frac{6.97\times10^{6}}{\tau_{in}}\frac{dR}{R^{2}} = \int_{R_{in}}^{\lambda R_{in}}\frac{1.65\times10^{7}\eta}{[(0.09/\tau^{1.5}) + 0.66]}\frac{dR}{R^2}$.
\end{center}
As $\tau \gtrsim 1$ and $R_{ph} = R_{in}\tau_{in} = 2.17\times10^{11}$ cm,
\begin{eqnarray}
1.5\times10^{-3}(1 - {\rm ln}\lambda) + 3.21\times10^{-5}(1 - 1/\lambda) \nonumber \\
 = 1.15\times10^{-4}\eta(1 - 1/\lambda)\tau_{in}.
\end{eqnarray}
For $\lambda \sim 2$ i.e. for protons to cool down to electron energies in $t^{\prime} = 2 t_{dyn}^{\prime}$, $\eta \tau_{in} \sim 8.28$. This means that for $\tau_{in} \gtrsim 10$, the protons cool down too fast and super-Coulomb interaction cannot keep the electrons hot beyond $t^{\prime} = 2 t_{dyn}^{\prime}$. As a result, the photons will not be up-scattered to larger energies and the output spectrum will not have a high-energy non-thermal power-law tail as seen in the observed Band spectrum. This necessitates the heating of electrons by some alternate sub-photospheric dissipation mechanism such as internal shocks \citep{RM94,LB10} or magnetic reconnection events \citep{Gia06}. Even though the electrons tend to cool down rapidly due to Comptonization, their energy can still be maintained at $\gamma_e \gtrsim \gamma_{e,crit}$ provided the episodic heating events are frequent.

\section{Electron heating by repeated sub-photospheric dissipation events}  
\label{e_rh}
In this section, we compute the threshold electron energy $\gamma_{e,crit}$ in a more exact manner by including the effect of photon and electron cooling due to adiabatic expansion of the outflow. As the electrons are maintained at $\gamma_{e} \sim \gamma_{e,crit}$ by energy gain from either Coulomb collisions or repeated dissipation events and subsequent cooling due to IC, we can further constrain the injected energy and the number of episodic dissipation events required for the output photon spectrum to have a Band-like shape.
As earlier, it is reasonable to estimate the electron energy assuming IC is the dominant cooling process as $t_{IC} \ll t_{dyn}$.

\subsection{Electron energy in terms of $L_{\gamma}$ and $\tau_{in}$}
The threshold electron energy $\gamma_{e,crit}$ (discussed in Section 2.1) can now be obtained but in a more self-consistent manner by accounting for the energy loss of the photons due to adiabatic cooling. The photon energy reduces due to adiabatic loss by the factor $(R_{ph}/R_{in})^{-2/3} \sim \tau_{in}^{-2/3}$ until they escape the photosphere (see Section 4).\\
{\it Total energy gained by the photons = total energy transferred by the electrons through Comptonization}
\begin{eqnarray}
N_{\gamma}\tau_{in}^{2/3}\left(E^{\prime}_{\gamma,avg,obs} - E^{\prime}_{\gamma,avg,i}\right) \nonumber \\
= N_{e}\int_{0}^{\tau_{in}t_{dyn}^{\prime}}(4/3)U_{\gamma}^{\prime}\tau_{in}^{-2/3}\sigma_{T}\gamma_{e}^{2}\beta_{e}^{2}c\ dt^{\prime}.\nonumber
\end{eqnarray}
Rewriting as an R-integral with $R = R_{in} + \beta c\Gamma t^{\prime}$ and using, $E^{\prime}_{\gamma,avg,i} \ll E^{\prime}_{\gamma,avg,obs}$ and $L_{\gamma}/L \sim E_{\gamma}/E \sim 3.2\times10^{-2}$,
\begin{eqnarray}
10^{5}\tau_{in}^{2/3}E^{\prime}_{\gamma,avg,obs} = 8.72\times10^{6}L_{\gamma,50}\tau_{in}^{-2/3}\int_{R_{in}}^{\tau_{in}R_{in}}(\gamma_{e}^{2} - 1)\frac{dR}{R^2} \nonumber \\
\sim \frac{8.72\times10^{6}L_{\gamma,50}\tau_{in}^{-2/3}(\gamma_{e}^{2} - 1)}{R_{in}}.\nonumber
\end{eqnarray}
Substituting $R_{in} \sim 2.17\times10^{11}\ \tau_{in}^{-1}$ cm yields
\begin{equation}
\gamma_{e,crit} \sim \sqrt{1 + 1.2\ L_{\gamma,50}^{-1}\ \tau_{in}^{1/3}}.
\label{eqn1}
\end{equation}
The critical electron energy obtained here is similar to $\gamma_{e,crit} \sim 1.352$ obtained in Section 2.1 for small initial optical depths $\tau_{in} \sim 1$. However, the value of $\gamma_{e,crit}$ obtained from equation (\ref{eqn1}) can be considerably larger when $\tau_{in} \gtrsim 10$ as shown in Table 1, especially for smaller $L_{\gamma}$.

\subsection{The $E_{inj}-\tau_{in}$ correlation}
Electron-photon collisions by themselves cannot inject significant amount of energy into the electrons and keep them sufficiently hot such that the scattered photons have a Band-like output spectrum. The electrons need to be heated additionally by some alternate dissipation mechanism which can transfer considerable amount of energy to them. Here we consider repeated sub-photospheric dissipation events that can re-accelerate the electrons as well as protons to their initial energies. We constrain the energy injected per electron $E_{inj} = N_{rh}(\gamma_{e,in} - 1)m_e c^2$, using the fact that the electrons remain in equilibrium with energy $\gamma_{e,crit} \sim \sqrt{1 + 1.2\ L_{\gamma,50}^{-1}\ \tau_{in}^{1/3}}$, from these heating episodes and subsequent cooling due to IC and adiabatic expansion.\\
{\it Equilibrium energy of the electrons ($E_{e,crit}$) = energy gained by Coulomb collisions and repeated dissipation events ($\Delta E_{e,Coul} + \Delta E_{e,rh}$) - energy lost due to IC and adiabatic cooling ($\Delta E_{e,IC} + \Delta E_{e,ad}$)} 
\begin{eqnarray}
N_{e}\tau_{in}^{4/3}(\gamma_{e} - 1)m_{e}c^{2} = N_{e}\int_{0}^{\tau_{in}t_{dyn}^{\prime}}\frac{5\times10^{-19}n_{e}^{\prime}\beta_{p}^{2}\tau_{in}^{-4/3}}{[0.73(\gamma_{e} - 1/\gamma_{e})^{3/2} + \beta_{p}^{3}]} dt^{\prime}\nonumber \\
+ N_{e}N_{rh}(\gamma_{e,in} - 1)m_{e}c^{2} - N_{e}\int_{0}^{\tau_{in}t_{dyn}^{\prime}}(4/3)U_{\gamma}^{\prime}\tau_{in}^{-2/3}\sigma_{T}\gamma_{e}^{2}\beta_{e}^{2}c\ dt^{\prime},
\label{tau_Nrh1}
\end{eqnarray}
where $\gamma_e \approx \gamma_{e,crit}$ and $\tau_{in}^{-4/3}$/$\tau_{in}^{-2/3}$ is the adiabatic cooling factor for relativistic electrons/photons (see Section 4 for more details). Here we consider the episodic dissipation events to be equally spaced over the jet expansion timescale $\tau_{in}t_{dyn}^{\prime}$ and to supply fixed energy (equal to initial energy, $\gamma_{e,in}$/$\gamma_{p,in}$) to the electrons/protons at each instance.
Although the seed photons in our system are generated from the synchrotron emission of fast cooled electrons at $\tau_{in} \gtrsim 100$ (see equation \ref{ph_seed}), the associated synchrotron energy loss rate for these electrons can be effectively ignored in comparison to the IC cooling rate in equation (\ref{tau_Nrh1}) as the synchrotron power $P_{syn} = (U_B^{\prime}/U_{\gamma}^{\prime})P_{IC}$ is significantly smaller compared to the IC cooling rate $P_{IC}$ for $\tau_{in} \sim {\rm few}-10{\rm s}$ due to the rapidly decreasing field strength.

The average electron energy does not change appreciably with scatterings when $\gamma_e \sim 1 + 1/(8\tau_{in})$ after $N_{Comp}$ scatterings as the IC and Coulomb interaction timescales are similar. For protons with initial energy $\gamma_{p,in} \sim 2$ and cooling adiabatically, $\gamma_{p} \sim 1 + 1/\tau_{in}$ i.e., $\beta_{p}^{2} = 1 - (1 + 1/\tau_{in})^{-2} \approx 2/\tau_{in}$. Substituting $n_{e}^{\prime}$, $\beta_{p}^{2}$, $U_{\gamma}^{\prime}$ and rewriting equation (\ref{tau_Nrh1}) as an integral over $R$,
\begin{eqnarray}
\tau_{in}^{4/3}(\gamma_{e} - 1)m_{e}c^{2} = \frac{1.09\times10^7 \times (2/\tau_{in}) \tau_{in}^{-4/3}}{[0.73(\gamma_{e} - 1/\gamma_{e})^{3/2} + (2/\tau_{in})^{3/2}]} \int_{R_{in}}^{\tau_{in}R_{in}}\frac{dR}{R^2} \nonumber \\ + N_{rh}(\gamma_{e,in} - 1)m_{e}c^{2} - \frac{6.97\times10^6 \tau_{in}^{-2/3}}{\tau_{in}}\int_{R_{in}}^{\tau_{in}R_{in}}\frac{dR}{R^2}.\nonumber
\end{eqnarray}
Further simplification and substituting $R_{in} = 2.17\times10^{11}\ \tau_{in}^{-1}\ {\rm cm}$ yields
\begin{eqnarray}
\tau_{in}^{4/3}(\gamma_e - 1)m_e c^2 = \frac{10^{-4}\tau_{in}^{-4/3}}{[0.73(\gamma_{e} - 1/\gamma_{e})^{3/2} + (2/\tau_{in})^{3/2}]} \nonumber \\
+ E_{inj,cr}(\tau_{in}) - 3.21\times 10^{-5}\tau_{in}^{-2/3},
\label{Nrh_ad}
\end{eqnarray}
which constrains the critical injected energy $E_{inj,cr}(\tau_{in}) = N_{rh}(\gamma_{e,in} - 1)m_{e}c^{2}$ per electron in terms of $\tau_{in}$. 
It should be noted that equation (\ref{Nrh_ad}) is only a necessary and not sufficient condition to obtain Band-like photon spectrum as it determines the average photon energy but does not impose any constraints on the general shape of the photon spectrum. There exists a critical balance between the injected energy $E_{inj,cr}$ and the initial optical depth of the outflow $\tau_{in}$: for large $E_{inj}$, the photon peak energy $E_{\gamma,peak} \gg E_{\gamma,obs} \sim 300\ {\rm keV}$ while $E_{\gamma,peak} \ll E_{\gamma,obs}$ for large $\tau_{in}$, due to significant energy loss from adiabatic cooling.
It should be noted that $n_{e}^{\prime} \approx 10^{16}\ {\rm cm^{-3}}$ and $R \approx 10^{11}\ {\rm cm}$ implies that the total injected energy $E_{inj,tot} = N_e E_{inj,cr}$ can be significantly larger for a typical GRB fireball as the electron number $N_e \approx N_p \sim 10^{50-51}$.

\section{Photospheric code description}
\label{code_desc}
In this section, we describe the basic structure of our MCRaT code and the associated physics. We list the jet parameters along with the initial position, energy and velocity distributions of the particles (electrons and protons) and the photons. We then discuss how the particles and photons in the jet are affected by the physical processes such as adiabatic cooling, Coulomb, IC and pair production/annihilation. Next, we briefly describe the algorithm of our photospheric MCRaT code.

\subsection{Jet parameters}
The jet parameters used as input for the MCRaT code are:
\begin{itemize}
\item{{\it Isotropic equivalent luminosity of the outflow,} $L$}: The bulk of the jet luminosity is contributed by the protons as they have most of the jet kinetic energy. We consider $L = 10^{51},\ 10^{52}\ {\rm erg/s}$ \citep{Liang07,WP10}.

\item{{\it Jet bulk Lorentz factor,} $\Gamma$}: The bulk Lorentz factor is related to $L$ and the peak photon energy $E_{\gamma,peak}$. We consider $\Gamma = 30, 100, 300$ in this work \citep{Xue09,Liang10}.
For small $\Gamma \lesssim 30$, the outflow might not produce a GRB successfully and rather show up as a X-ray rich GRB or a X-ray flash with a different spectrum \citep{Huang02}.

\item{{\it Initial optical depth,} $\tau_{in}$}: The optical depth $\tau$ is measured in relation to $R = L\sigma_T/(8\pi m_p c^3 \beta \Gamma^3 \tau)$ which is the radial distance of a photon from the central engine in the observer frame. $\tau_{in}$ corresponds to the radial distance from the central engine where all the particles and photons are injected and $\tau = 1$ denotes the photospheric radius where all the photons escape. Here we consider $\tau_{in} = 10, 20, 40$.
\end{itemize}

\subsection{Particles and their distributions}
\label{Par_dist}
Now we describe the initial energy and velocity distributions of the electrons, protons and photons in the jet. 
\begin{itemize}
\item{{\it Electrons and protons:}} We consider a charge-neutral jet with particle number $N_{e} = N_{p} = 2\times 10^{2}$ \citep{CL15,MB18}. We show that $2\times10^{2}$ electrons are sufficient in order to represent the outflow. The initial velocities of all the electrons and protons are distributed randomly in the jet-comoving frame (see Appendix B1 of \citealt{Santana16}). All the particles are uniformly distributed in the jet-comoving frame at initial time $t=0$. The initial energy of the electrons are determined from the Maxwellian distribution with temperature $k_{B}T_{e,in}^{\prime} = (\gamma_{e,ad,in}-1)(\gamma_{e,in} - 1)m_e c^2$ while the protons are mono-energetic with $\gamma_{p} = \gamma_{p,in}$. We consider $\gamma_{e,in} = 2, 10, 30, 100$ and $\gamma_{p,in} = 1.01,1.1$ for our simulations. 

\item{{\it Photons:}} In order to maintain $N_{\gamma}/N_{e} = 10^{5}$, we consider $N_{\gamma} = 2\times10^{7}$ for our simulations \citep{MB18}. The initial velocities of the photons are randomly distributed in the comoving frame of the jet and the photon positions are uniformly distributed within a cone with solid angle $1/\Gamma$ in the direction of the observer. The initial photon energy distribution is given by the synchrotron distribution for fast cooling electrons \citep{Granot00,Piran04}
\begin{equation}
\label{ph_seed}
f_{\nu} = \left\{
\begin{array}{ll}
\left(\frac{\nu_{ac}}{\nu_{sa}}\right)^{11/8}\:\left(\frac{\nu}{\nu_{ac}}\right)^{2}, & \nu_{min} < \nu < \nu_{ac}\\
\left(\frac{\nu}{\nu_{sa}}\right)^{11/8}, & \nu_{ac} < \nu < \nu_{sa} \\
\left(\frac{\nu}{\nu_{sa}}\right)^{-1/2}, & \nu_{sa} < \nu < \nu_{m} \\
\left(\frac{\nu_{m}}{\nu_{sa}}\right)^{-1/2}\:\left(\frac{\nu}{\nu_{m}}\right)^{-p/2}, & \nu_{m} < \nu < \nu_{max}\\
\end{array}
\right. 
\end{equation}
where $f_{\nu}$ is the peak normalised photon flux per unit frequency and $p = 2.5$ is the spectral index at high energies (\citealt{KZ15}). 

For typical GRB parameters: $\epsilon_{B} = 0.1$, $\epsilon_{e} = 0.1$, number of peaks in the burst $N_{GRB} = 10^{2}$ and duration of the burst $T_{GRB} = 10$ s (see, \citealt{Granot00}), we have $E_{\gamma,peak}^{\prime} = h\nu_{sa}^{\prime} = 2\ {\rm eV}$, $h\nu_{min}^{\prime} = 1.5\times10^{-9}\ E_{\gamma,peak}^{\prime}$, $h\nu_{ac}^{\prime} = 10^{-2}\ E_{\gamma,peak}^{\prime}$, $h\nu_{m}^{\prime} = 500\ E_{\gamma,peak}^{\prime}$ and $h\nu_{max}^{\prime} = 1.5\times10^{4}\ E_{\gamma,peak}^{\prime}$. 
\end{itemize}

\subsection{Physical processes in the outflow}
Here we discuss the interactions between the electrons, protons and photons which can further affect the output photon spectrum. The physics of the relativistic outflow can be broadly decoupled into four categories:
\begin{itemize}
\item{{\it Adiabatic cooling:}} As the relativistic jet expands outward, the energies of the electrons, protons and photons reduce considerably due to adiabatic cooling. The energies are affected by adiabatic cooling as 
\begin{eqnarray}
(\gamma_{e,f} - 1)/(\gamma_{e,i} - 1) = \left(R_{in} + \beta c \Gamma t^{\prime}_{f}/R_{in} + \beta c \Gamma t^{\prime}_{i}\right)^{-2(\gamma_{ad,e} - 1)},\nonumber \\
(\gamma_{p,f} - 1)/(\gamma_{p,i} - 1) = \left(R_{in} + \beta c \Gamma t^{\prime}_{f}/R_{in} + \beta c \Gamma t^{\prime}_{i}\right)^{-2(\gamma_{ad,p} - 1)},\nonumber \\
E_{\gamma,f}/E_{\gamma,i} = \left(R_{in} + \beta c \Gamma t^{\prime}_{f}/R_{in} + \beta c \Gamma t^{\prime}_{i}\right)^{-2/3},
\end{eqnarray}
where the subscript $i/f$ denotes the initial/final value of the physical quantity and $\gamma_{ad,e/p} = (4\gamma_{e/p} +1)/(3\gamma_{e/p})$ is the electron/proton adiabatic index. This energy scaling with $R$ is valid as the radial width of the jet is fixed and the electron density $n_{e}^{\prime}$ decreases as $R^{-2}$.

\item{{\it Coulomb collisions:}} The electrons are continuously heated by the protons that carry most of the energy in the jet. The electrons also interact with each other to quickly attain thermal equilibrium that is given by a quasi-Maxwellian distribution after every energy transfer event. The proton-electron and electron-electron energy transfer rates are \citep{Schlickheiser02} 
\begin{eqnarray}
\dot{E}_{e-p} = \frac{5\times10^{-19}n_{e}^{\prime}\beta_{p,avg}^{2}}{8.3\times10^{-15}T_{e,avg}^{\prime 3/2} + \beta_{p,avg}^{3}},\nonumber \\
\dot{E}_{e-e} = \frac{5\times10^{-19}n_{e}^{\prime}\beta_{e,avg}^{2}}{8.3\times10^{-15}T_{e,avg}^{\prime 3/2} + \beta_{e,avg}^{3}},
\label{Coul}
\end{eqnarray}
where $\beta_{p,avg}$, $\beta_{e,avg}$ and $T_{e,avg}^{\prime}$ are number-averaged quantities. This is valid as the electrons undergo Coulomb interaction with the average proton/electron distribution around them and vice-versa. The expression for electron-electron energy transfer rate $\dot{E}_{e-e}$ is almost the same as the proton-electron energy transfer rate $\dot{E}_{e-p}$ except that $\beta_{p,avg}$ is replaced by $\beta_{e,avg}$ as the nature of the underlying interaction is essentially the same. 

The electron distribution is re-initialized to Maxwellian distribution on a timescale $t_{e-e}^{\prime} = (\gamma_{e,avg}-1)m_e c^2/\dot{E}_{e-e} \ll t_{e-p}^{\prime}$. It should be noted that $T_{e,avg}^{\prime}$ in equation (\ref{Coul}) may not always correspond to a Maxwellian distribution with peak energy $\gamma_{e,avg}$, especially for large $N_{\gamma}/N_{e} \sim 10^{5}$. However, the electron-electron and electron-proton interactions are still described by equation (\ref{Coul}) provided $t_{e-e}^{\prime}$ is comparable to $t_{IC}^{\prime}$ such that the electrons attain a quasi-thermal distribution very rapidly.

\item{{\it IC scattering:}} As the average photon energy is much smaller than that of the electrons, the photons continue to scatter off of the electrons and gain energy until either $E_{\gamma,avg}^{\prime} \approx (\gamma_{e,avg} - 1)m_e c^2$ or $\tau = 1$. The distance $s^{\prime}$ that a photon travels before scattering an electron is given by the probability density $p(s^{\prime}) \propto {\rm exp}(-s^{\prime}/l_{mfp}^{\prime})$, where $l_{mfp}^{\prime} = 1/(n_{e}^{\prime}\sigma_T)$ is the photon mean free path. Not all the $N_{e}$ electrons in the jet are equally likely to scatter the photon and the probability of scattering for a particular electron with a photon is (see \citealt{MB18}, for details)
\begin{equation}
P_{sc}(\beta_{e},\theta_{e}^{\prime}) = \frac{1}{4\pi \beta_e^{2}}(1 - \beta_{e} {\rm cos}\ \theta_{e}^{\prime}),
\label{Psc}
\end{equation} 
where $\beta_{e}$ is the electron speed divided by the speed of light and $\theta_{e}^{\prime}$ is the angle between the electron and photon velocities before scattering occurs. The average number of scatterings that a photon experiences before it escapes the photosphere is $\sim 2\tau_{in}$ \citep{Begue13}.

\item{{\it Pair production/annihilation:}} Due to the episodic global dissipation events in the jet, the electrons are often accelerated to highly relativistic energies with $\gamma_{e} = \gamma_{e,in} \sim 100$ and can then scatter energetic photons with $E_{\gamma}^{\prime} \gtrsim 10\ E_{\gamma,peak}^{\prime}$ to energies $\gtrsim 4 E_{\gamma,avg}^{\prime}\gamma_{e}^{2} \sim m_e c^2 \sim 5\times10^{5}\ {\rm eV}$, before cooling down rapidly to non-relativistic energies. For photons with a fast-cooled synchrotron spectrum that we consider (see equation \ref{ph_seed}), a considerable fraction $\sim 30\%$ have sufficient energy to generate electron-positron pairs and thereby increase/decrease the electron/photon number in the jet considerably. This can affect the output photon spectrum significantly if the number of pairs created $N_{e^{-}e^{+}}$ is comparable to $N_{e}$, by altering the photon to electron number ratio $N_{\gamma}/N_{e}$, especially for large $\tau_{in} \gtrsim 10$ (see Appendix \ref{Appendix_pair}, for more details).
\end{itemize} 

\subsection{MCRaT code description}
The photospheric MCRaT code is described in significant detail in \cite{MB18}, but here we describe it briefly. Initially, travel distances (distance that each photon travels before scattering an electron) are drawn for all photons depending on their mean free path and the photons are propagated. The new positions of the photons in the lab frame are evaluated to check if any photon escapes the photosphere, in which case the energy in the lab frame is calculated and stored. All other photons are stored in a priority queue where the photons are ordered by increasing values of travel distances. In the next step, the photon at the top of the queue is propagated, a proton is randomly selected and an electron is selected using the scattering probability, $P_{sc}$. The energies of the particles and photons are then updated due to adiabatic cooling and Coulomb collisions. Next, the outgoing velocities and energies of the photon and the electron are calculated if IC scattering occurs, provided the photon energy dependent scattering cross section is sufficiently large.

\begin{figure*}
    \begin{subfigure}[tp]{0.49\linewidth}
    \centering
    \includegraphics[width=0.97\linewidth]{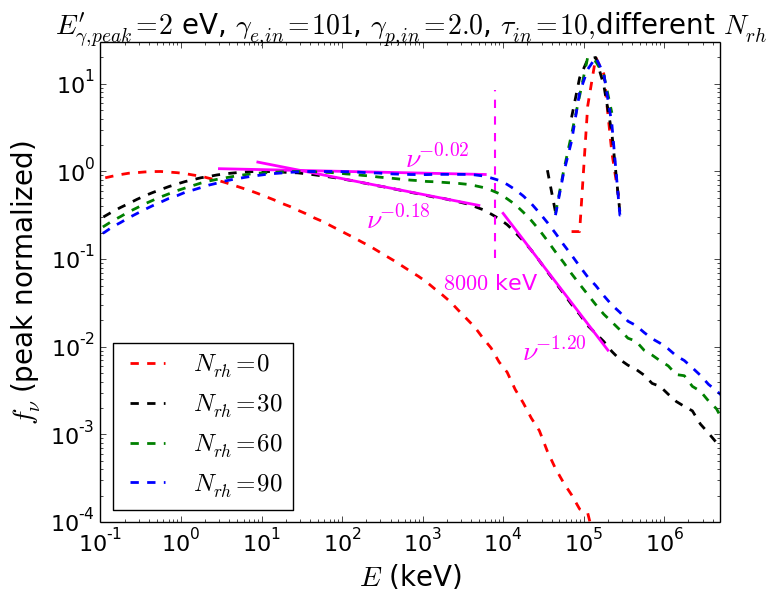} 
    %\includegraphics[width=0.97\linewidth]{1.1.ps}
    %\caption{Initial condition} 
    %\label{fig1:a} 
    %\vspace{2ex}
  \end{subfigure}%% 
  \begin{subfigure}[tp]{0.49\linewidth}
    \centering
    \includegraphics[width=0.90\linewidth]{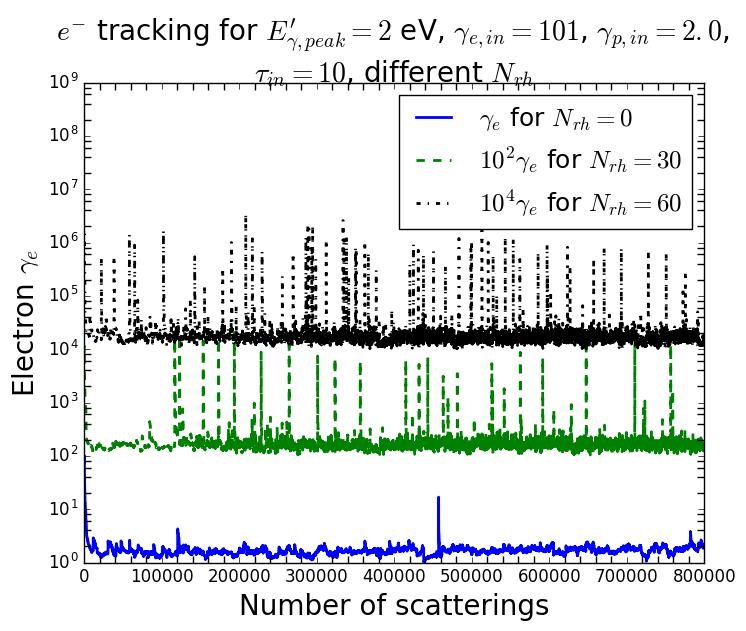}  
    %\includegraphics[width=0.97\linewidth]{1.2.ps}
    %\caption{Rupture} 
    %\label{fig7:b} 
    %\vspace{2ex}
  \end{subfigure} \\ 
  \begin{subfigure}[tp]{0.49\linewidth}
    \centering
    \includegraphics[width=0.97\linewidth]{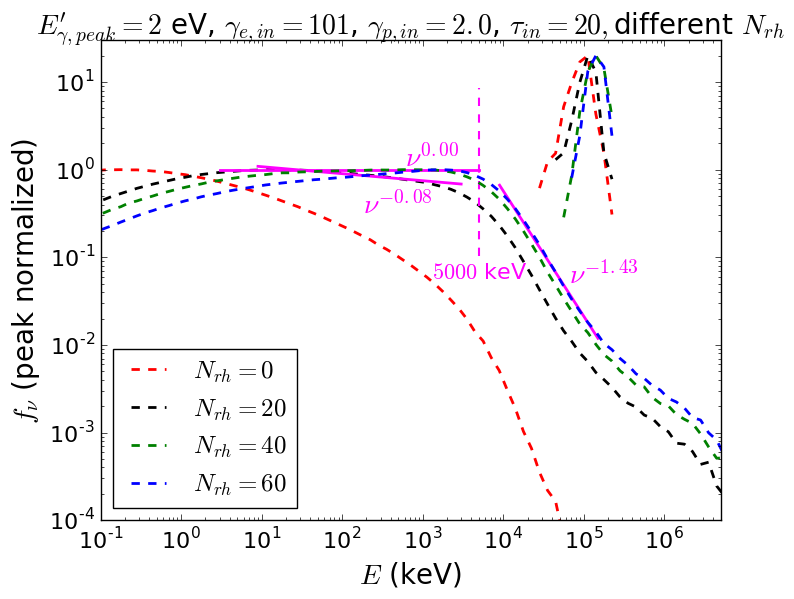} 
    %\includegraphics[width=0.97\linewidth]{1.2.ps}
    %\caption{Rupture} 
    %\label{fig7:b} 
    %\vspace{2ex}
  \end{subfigure} 
    \begin{subfigure}[tp]{0.49\linewidth}
    \centering
    \includegraphics[width=0.90\linewidth]{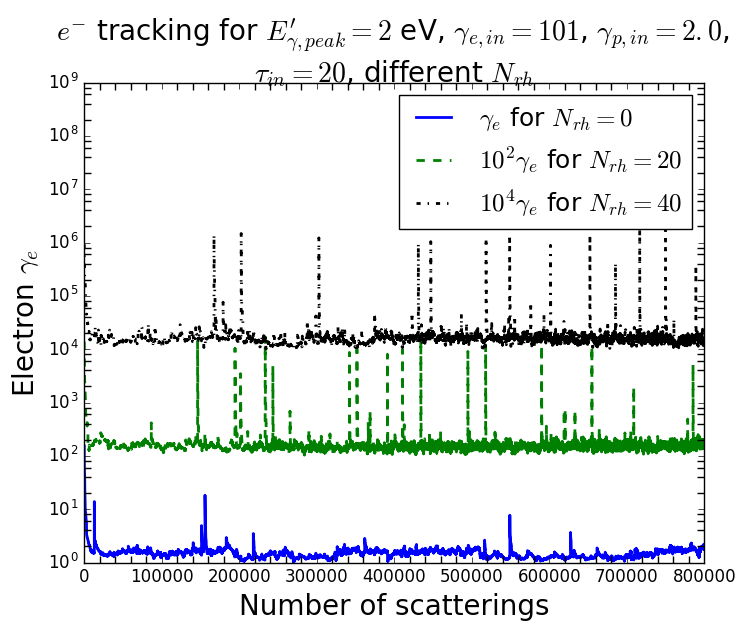} 
    %\includegraphics[width=0.97\linewidth]{1.1.ps}
    %\caption{Initial condition} 
    %\label{fig1:a} 
    %\vspace{2ex}
  \end{subfigure} \\%% 
  \begin{subfigure}[tp]{0.49\linewidth}
    \centering
    \includegraphics[width=0.97\linewidth]{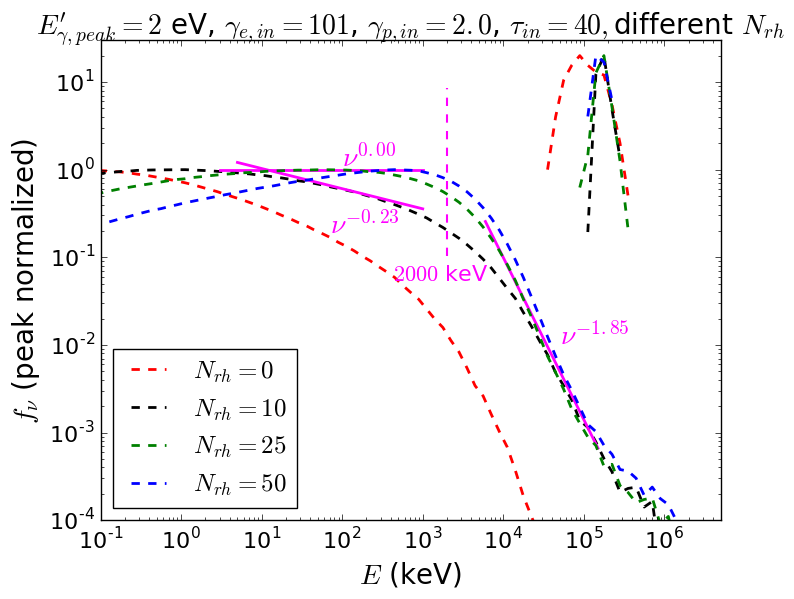} 
    %\includegraphics[width=0.97\linewidth]{1.2.ps}
    %\caption{Rupture} 
    %\label{fig7:b} 
    %\vspace{2ex}
  \end{subfigure} 
  \begin{subfigure}[tp]{0.49\linewidth}
    \centering
    \includegraphics[width=0.90\linewidth]{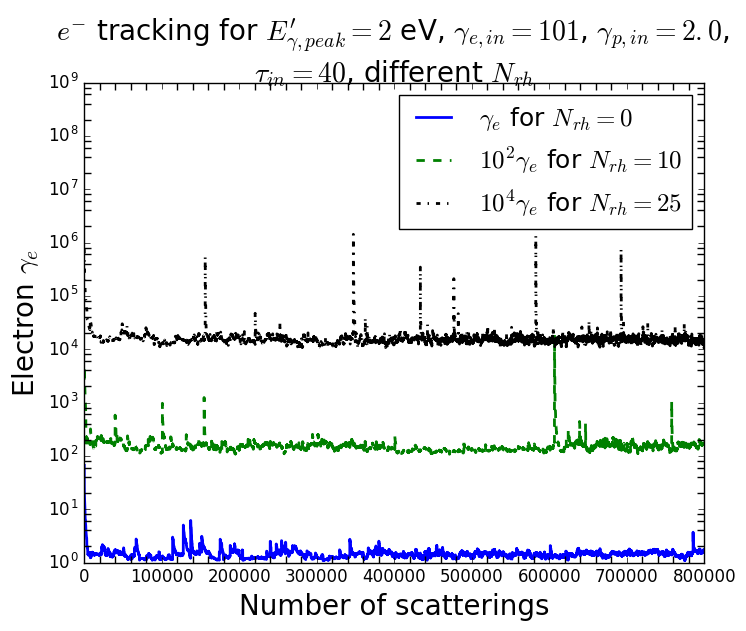} 
    %\includegraphics[width=0.97\linewidth]{1.2.ps}
    %\caption{Rupture} 
    %\label{fig7:b} 
    %\vspace{2ex}
  \end{subfigure} 
  \caption{
  {\it MCRaT simulations showing the effect of $N_{rh}$ at constant $\gamma_{e,in}$ and for different $\tau_{in}$, with input parameters:} $E_{\gamma,peak}^{\prime}=2$ eV, $\gamma_{e,in}=101$, $\gamma_{p,in}=2$, $L=10^{52}\ {\rm erg/s}$ and $\Gamma=300$. 
The left-half panels show the output photon spectra for different $\tau_{in}$ values -   
	{\it Top-left:} $\tau_{in} = 10$ and $N_{rh} = 0/30/60/90$, 
	{\it Center-left:} $\tau_{in} = 20$ and $N_{rh} = 0/20/40/60$,
	{\it Bottom-left:}  $\tau_{in} = 40$ and $N_{rh} = 0/10/25/50$. 
The right-half panels show the evolution of electron energy over multiple scattering events for the corresponding cases - 	 
	{\it Top-right:} $\tau_{in} = 10$ and $N_{rh} = 0/30/60$, 
	{\it Center-right:}  $\tau_{in} = 20$ and $N_{rh} = 0/20/40$,
	{\it Bottom-right:}  $\tau_{in} = 40$ and $N_{rh} = 0/10/25$.
The electron kinetic energy spectra at the end of each simulation are also shown in the left-half panels. 
The right-half panels show the energy evolution with scattering events for three different electrons that are selected randomly from our sample. The spikes in $\gamma_e$ correspond to energy injection/proton Coulomb collision/photon Comptonization events each resulting in a large energy transfer to the electron.%\\ 
 }
  \label{fig1} 
\end{figure*}

Then the next photon in the queue is drawn and electron-positron pair production cross section is evaluated. If the cross section is large, a new electron and positron are generated and the photons are not placed back in the queue. If the positron number is non-zero, a positron is drawn randomly and the pair annihilation cross section with the electron is calculated. Two new photons are created and added to the queue if the cross section is significant. Again, as initially, the photon at the top of the queue is propagated with its travel distance to check if $R \geq R_{ph}$. The photon is collected as a part of the observed spectrum if it manages to escape the photosphere, otherwise the method described above is repeated until a third of the total photons in the jet escape and a time-averaged output photon spectrum is obtained. In our MCRaT code, electron-photon scattering events are performed one at a time and the particles are re-accelerated to their initial distributions by dissipation events that are evenly spaced within scattering events.

\section{Photospheric simulation results}
\label{sim_results}

\begin{figure*}
  \begin{subfigure}[b]{0.5\linewidth}
    \centering
    \includegraphics[width=0.95\linewidth]{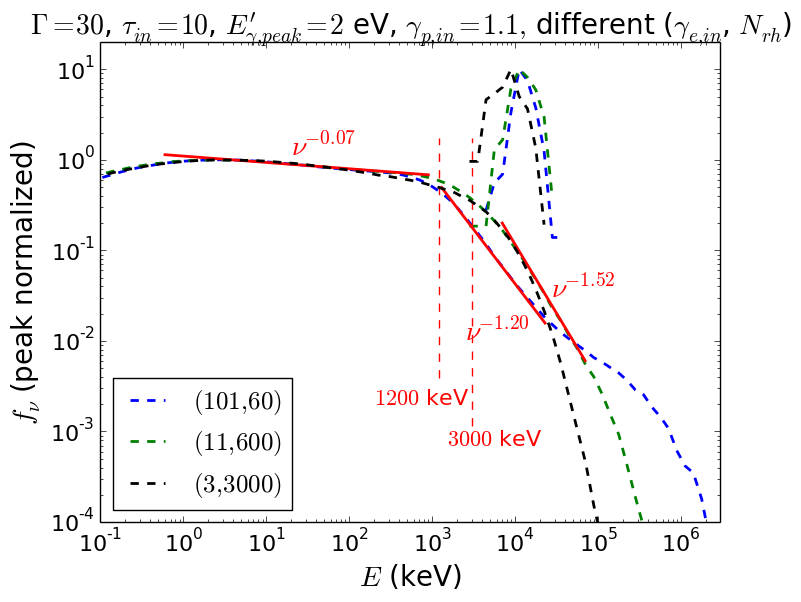} 
    %\includegraphics[width=0.97\linewidth]{3.1.ps}
    %\caption{Initial condition} 
    %\label{fig3:a} 
    %\vspace{1ex}
  \end{subfigure}%%
  \begin{subfigure}[b]{0.5\linewidth}
    \centering
    \includegraphics[width=0.95\linewidth]{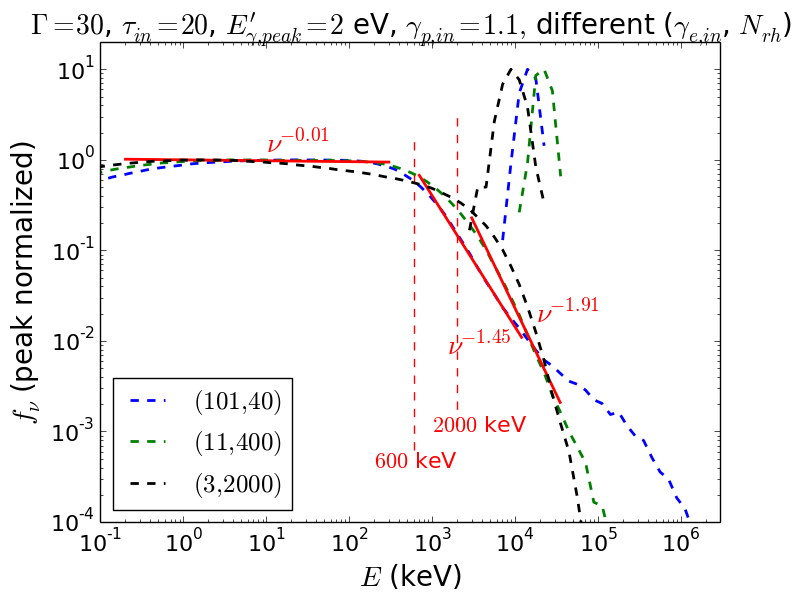} 
    %\includegraphics[width=0.97\linewidth]{3.2.ps}
    %\caption{initial condition} 
    %\label{fig3:b} 
    %\vspace{1ex}
  \end{subfigure}\\ 
   \begin{subfigure}[b]{0.5\linewidth}
    \centering
    \includegraphics[width=0.95\linewidth]{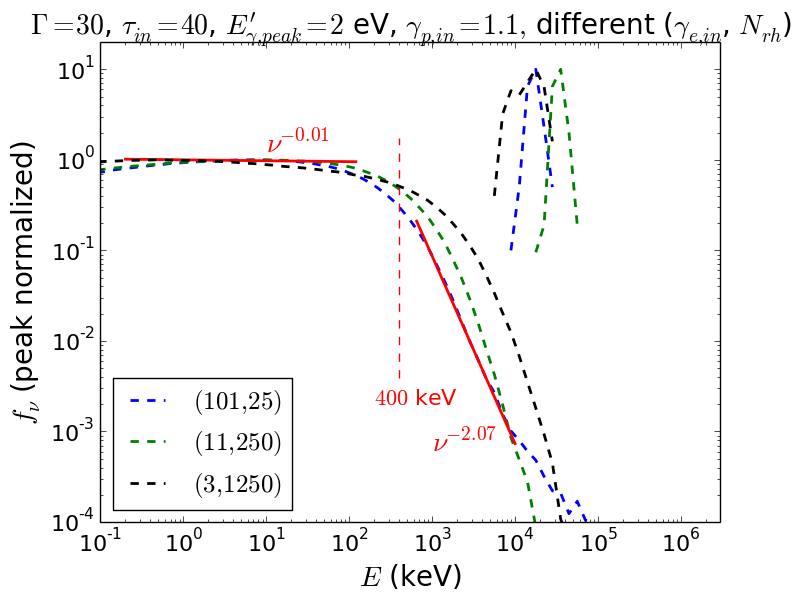} 
    %\includegraphics[width=0.97\linewidth]{3.1.ps}
    %\caption{Initial condition} 
    %\label{fig3:a} 
    %\vspace{1ex}
  \end{subfigure}\\%%
  \caption{{\it MCRaT simulations showing the effect of $\gamma_{e,in}$ at constant $E_{inj}=E_{inj,cr}(\tau_{in})$ and for different $\tau_{in}$, with input parameters:} $E_{\gamma,peak}^{\prime}=2$ eV, $\gamma_{p,in}=1.1$, $L=10^{52}\ {\rm erg/s}$ and $\Gamma=30$. 
         {\it Top-left panel:} $\tau_{in}=10$, $E_{inj} = 6000\ m_e c^2$ and $(N_{rh},\gamma_{e,in}) = (60,101)/(600,11)/(3000,3)$.
	{\it Top-right panel:} $\tau_{in}=20$, $E_{inj} = 4000\ m_e c^2$ and $(N_{rh},\gamma_{e,in}) = (40,101)/(400,11)/(2000,3)$.
	{\it Bottom panel:} $\tau_{in}=40$, $E_{inj} = 2500\ m_e c^2$ and $(N_{rh},\gamma_{e,in}) = (25,101)/(250,11)/(1250,3)$.	
	%{\it Top right: $\Gamma=30$, $\tau_{in}=20$:} 
	%$\gamma_{e,in}=100$ and $N_{rh}=(20,40,60)$, $\gamma_{e,in}=10$ and $N_{rh}=(200,400,600)$. \\
	%{\it Middle left: $\Gamma=30$, $\tau_{in}=40$:} 
	%$\gamma_{e,in}=100$ and $N_{rh}=(5,25,45)$, $\gamma_{e,in}=10$ and $N_{rh}=(50,250,450)$.
	%{\it Middle right: $\Gamma=100$, $\tau_{in}=10$:} 
	%$\gamma_{e,in}=10$ and $N_{rh}=(400,600,800)$, $\gamma_{e,in}=2$ and $N_{rh}=(2000,3000,4000)$. \\
	%{\it Bottom left: $\Gamma=100$, $\tau_{in}=20$:} 
	%$\gamma_{e,in}=10$ and $N_{rh}=(200,400,600)$, $\gamma_{e,in}=2$ and $N_{rh}=(1000,2000,3000)$.
	%{\it Bottom right: $\Gamma=100$, $\tau_{in}=40$:} 
	%$\gamma_{e,in}=10$ and $N_{rh}=(150,250,350)$, $\gamma_{e,in}=2$ and $N_{rh}=(750,1250,1750)$. \\
	}
  \label{fig2} 
\end{figure*}

In this section, we present the results of our photospheric MCRaT simulations. 
The photon energy spectrum and the electron kinetic energy spectrum are shown in the lab frame at the end of each simulation in all the figures.
We test our code by performing code validation tests which we describe briefly here (see \citealt{MB18}, for more details). First, we obtain the equilibrium distribution for Blackbody photons with energy $E_{\gamma,in}^{\prime} = 1000$ eV scattering Maxwellian electrons with $\gamma_{e,in}^{\prime}=1.001$. The equilibrium distribution at $\tau_{in} \sim 500$ for photons/electrons has energy dependence $f_{\nu} \propto \nu^{3}/f_{\nu} \propto \nu^{2}$ at low energies and $f_{\nu} \propto e^{-\nu}$ at high energies (see left panel of Fig. 1 in \citealt{MB18}). This is in very good agreement with the theoretical prediction that the equilibrium distribution of photons interacting with Maxwellian electrons at fixed energy approaches Bose-Einstein distribution with non-zero chemical potential. Next, we perform MCRaT simulations with the same input parameters as in Fig. 1 of \citet{CL15} for two different initial optical depths $\tau_{in}=5,75$ and obtain good agreement with their results for both the simulations (see \citealt{MB18}), which implies that our MCRaT code is working as expected. The photon and electron energy spectrum are Doppler boosted from the jet-comoving frame to the lab frame in all the figures. 
The electron kinetic energy spectra are peaked at significantly larger energies compared to the photon spectra for all the simulations as shown in the figures.
In the rest of this paper, we denote the low/high energy photon spectral index by $\alpha$/$\beta$ and the observed photon peak energy by $E_{\gamma,obs}$. 

In Figure \ref{fig1}, we present the simulation results with fixed $\gamma_{e,in}$ for four different values of $N_{rh}$ and $\tau_{in}=10,20,40$. The photon energy spectra are shown in the left-half panels while the electron energies are tracked over scattering events for the corresponding simulations and are shown in the right-half panels. The seed photons/electrons/protons have peak energies $E_{\gamma,peak}^{\prime}=2\ {\rm eV}$/$\gamma_{e,in}=101$/$\gamma_{p,in}=2.0$ with $L=10^{52}\ {\rm erg/s}$ and $\Gamma =300$ for all these simulations. We find that for $N_{rh}=0$, the photons in the output spectrum have energy $E_{\gamma,avg} \ll E_{\gamma,obs} \sim 1\ {\rm MeV}$ with a significantly harder high energy power-law tail $f_{\nu} \propto \nu^{-0.5}$ for all three $\tau_{in}$ considered (see, also, \citealt{MB18}). This is due to the fact that the electrons attain non-relativistic energies $\gamma_{e,Comp} \ll \gamma_{e,crit}$ very quickly ($N_{Comp} \sim 10^{4}$ in time $t_{Comp} \sim 10^{-2} R/\Gamma c \ll t_{dyn}$) in the absence of repeated dissipation events and cannot scatter photons to $\sim{\rm MeV}$ energies anymore. As $N_{rh}$ increases, the fraction of photons with $E_{\gamma} \gtrsim 1\ {\rm MeV}$ increases significantly and the output photon spectrum shows a distinct high energy power-law dependence.
We show the energy evolution over the entire scattering history for three electrons that are chosen randomly among $N_e = 200$ electrons in the jet. As opposed to the left-half panels that show the electron energy spectra at the end of each simulation, the right-half panels show the electron energy tracked after each scattering event. As $N_{\gamma}/N_e = 10^5$ and the average number of scatterings per photon is $\sim 2\tau_{in}$, the average number of scatterings per electron is $\sim 2\tau_{in}(N_{\gamma}/N_e) \sim 10^{6-7}$. We find that the electrons spend most of their time at non-relativistic energies except when energy injection/proton Coulomb collision/photon Comptonization events occur which accelerate them to relativistic energies. However, after each such event the electron again cools down rapidly to non-relativistic energy once it transfers almost all its excess kinetic energy to scatter a photon to $\sim$MeV energies.

For considerably larger values of $N_{rh}$, the photon spectrum peaks around $1-10\ {\rm MeV}$, which is expected as the hot electrons with $\gamma_{e} \gtrsim \gamma_{e,crit}$ can readily transfer their energy to the photons. We find that $\beta = -1.20/-1.43/-1.85$ and $E_{\gamma,peak} = 8/5/2 \ {\rm MeV}$ depend only on $\tau_{in} = 10/20/40$ and are roughly independent of $N_{rh}$. $E_{\gamma,peak}$ decreases whereas the high energy spectrum becomes steeper as $\tau_{in}$ increases, which is due to significant energy loss from adiabatic cooling. $E_{\gamma,peak}$ is larger than $E_{\gamma,obs}$ by a factor of $\sim 10$ even for large $\tau_{in}$, suggesting excess energy transfer to the photons either due to large $\Gamma$ or $\gamma_{e,in}$. It should be noted that $\alpha$ increases with injected energy $E_{inj} = N_{rh}(\gamma_{e,in}-1)m_e c^2$ and $\alpha \sim \alpha_{obs} \sim 0$ is obtained only for some critical energy $E_{inj,cr}(\tau_{in})$ as predicted by theory (see equation \ref{Nrh_ad}). From the simulation results, we obtain $E_{inj,crit}=6000/4000/2500\ m_e c^2$ for $\tau_{in}= 10/20/40$. The photon spectrum deviates from the observed Band spectrum both for large $E_{inj} \gg E_{inj,cr}$ (as $E_{\gamma,peak} \gg E_{obs}$) and large $\tau_{in} \gtrsim 50$ (as $E_{\gamma,peak} \ll E_{obs}$ and $|\beta| > |\beta_{obs}|$) due to adiabatic energy loss and geometrical broadening effects \citep{ST80,Poz83}.

In Figure \ref{fig2}, we present the simulation results for fixed injected energies $E_{inj,cr} = N_{rh,cr}(\gamma_{e,in}-1)\ m_e c^2 = 6000/4000/2500\ {\rm m_e c^2}$ at $\tau_{in}=10/20/40$ and different $\gamma_{e,in}=3,11,101$. The photons/protons are initialized with $E_{\gamma,peak}^{\prime}=2$ eV/$\gamma_{p,in}=1.1$ with jet parameters, $L=10^{52}\ {\rm erg/s}$ and $\Gamma=30$. We see that $\alpha \sim 0$ is practically unaffected by any decrease in electron energy $\gamma_{e,in}$ (irrespective of $\tau_{in}$) and is solely determined by the critical injected energy $E_{inj,cr}(\tau_{in})$. As $\gamma_{e,in}$ for a given $E_{inj,cr}$ increases, the photons tend to have lower peak energy $E_{\gamma,peak}$ and there are fewer/more photons with $E_{\gamma} \sim 1-10\ {\rm MeV}$/$\gtrsim 100\ {\rm MeV}$. This is expected as the electrons with $\gamma_{e,in} = 101$ are accelerated much less frequently compared to those with $\gamma_{e,in} \sim 3-11$ and then subsequently cool down very rapidly to non-relativistic $\gamma_{e}$ after being considerably hotter for a shorter duration $\sim 10^{-3}\ t_{dyn}$ when they accelerate many photons to $E_{\gamma} \gtrsim 100\ {\rm MeV}$. The high energy bump in $f_{\nu}$ and deviation from power-law behaviour for large $\gamma_{e,in}$ is seen only at moderate $\tau_{in} \lesssim 20$ and is not appreciable for larger $\tau_{in} \gtrsim 40$ as the high energy photons cool down rapidly from adiabatic losses. 

We find an increase in $|\beta|$ with decrease in $\gamma_{e,in}$ for a fixed $E_{inj,cr}(\tau_{in})$ as well as with increase in $\tau_{in}$. Moreover, $\beta \sim \beta_{obs}$ for $\gamma_{e,in} \sim {\rm few\ 10{\rm s}}-100$ and $\tau_{in} \lesssim 20$ while the high-energy spectrum is much steeper, $f_{\nu} \propto \nu^{-2}$ for $\tau_{in} \gtrsim 40$, almost independent of $N_{rh}$. The photon energy peak is much larger than $E_{\gamma,obs}$ especially for smaller $\tau_{in}$: $E_{\gamma,peak}/E_{obs} \sim 5-10/2-5/1$ for $\tau_{in} \sim 10/20/40$. While relatively continuous energy injection (small $\gamma_{e,in} \sim {\rm few}$ and large $N_{rh,cr} \sim {\rm few}\ 1000{\rm s}$) results in steeper high energy spectra $|\beta|>|\beta_{obs}|$ along with $E_{\gamma,peak}/E_{\gamma,obs} \gtrsim 10$, episodic energy injection (large $\gamma_{e,in}\sim 100$ and small $N_{rh,cr} \sim {\rm few}\ 10{\rm s}$) gives a high energy power-law spectrum consistent with observations for moderate optical depths $\tau_{in} \lesssim 20$. In order to have both $E_{\gamma,peak} \sim 500\ {\rm keV}$ and $|\beta| \sim 1.2-1.5$, the particles and photons have to be initialized at $\tau_{in} \sim 20-40$ and $E_{inj,cr}(\tau_{in}) \sim 2500-4000\ m_e c^2$ energy needs to be injected into electrons with $\gamma_{e,in} \sim {\rm few}\ 10{\rm s}$.

\begin{figure*}
  \begin{subfigure}[b]{0.5\linewidth}
    \centering
    \includegraphics[width=0.97\linewidth]{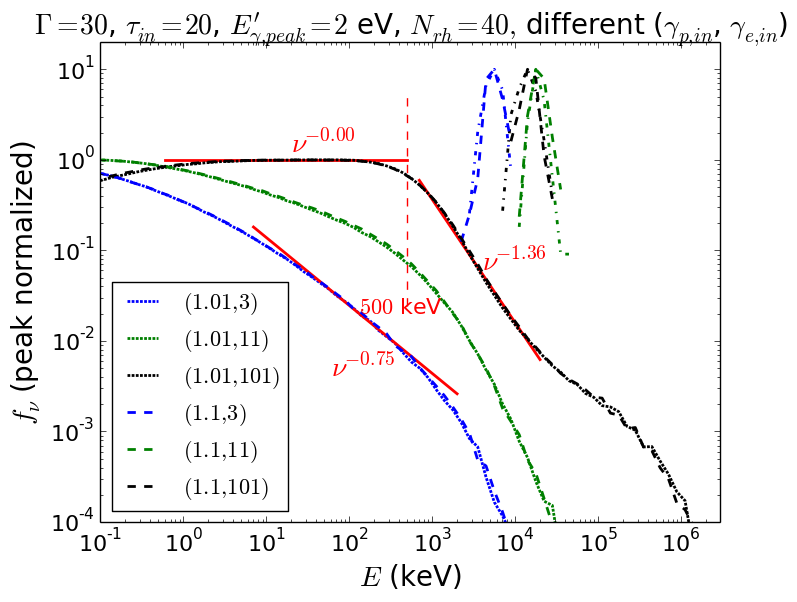}
    %\includegraphics[width=0.97\linewidth]{4.1.ps} 
    %\caption{Initial condition} 
    %\label{fig1:a} 
    %\vspace{1ex}
  \end{subfigure}%% 
  \begin{subfigure}[b]{0.5\linewidth}
    \centering
    \includegraphics[width=0.97\linewidth]{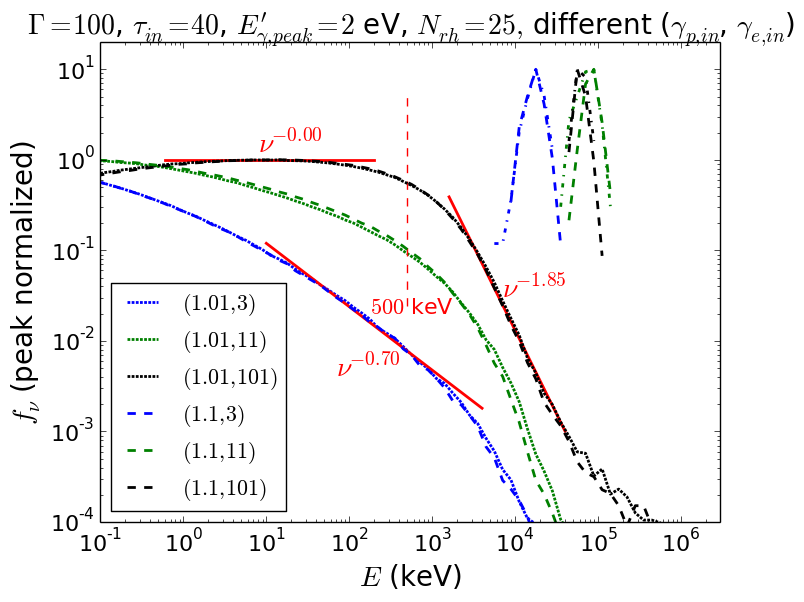}
    %\includegraphics[width=0.97\linewidth]{4.2.ps}  
    %\caption{Initial condition} 
    %\label{fig1:a} 
    %\vspace{1ex}
  \end{subfigure}%%  
  \caption{{\it MCRaT simulations showing the effect of $\gamma_{p,in}$ and $\gamma_{e,in}$ for $N_{rh}=40\ (\tau_{in}=20, \Gamma=30)$ and $N_{rh}=25\ (\tau_{in}=40, \Gamma=100)$, with input parameters:}
  $E_{\gamma,peak}^{\prime}=2$ eV and $L=10^{51}$ erg/s.	 
	{\it Left panel:} $\Gamma=30$, $\tau_{in}=20$, $N_{rh}=40$: 
	$\gamma_{p,in}=(1.01,1.1)$ and $\gamma_{e,in}=(3,11,101)$. 	
	{\it Right panel:} $\Gamma=100$, $\tau_{in}=40$, $N_{rh}=25$: 
	$\gamma_{p,in}=(1.01,1.1)$ and $\gamma_{e,in}=(3,11,101)$. %\\
 }
  \label{fig3} 
\end{figure*}

In Figure \ref{fig3}, we present the simulation results for fixed $N_{rh,cr}(\tau_{in})=40/25$ at $\tau_{in}=20\ (\Gamma=30)/40\ (\Gamma=100)$ with different combinations of $\gamma_{e,in}=3,11,101$ and $\gamma_{p,in}=1.01,1.1$. The seed photons have energy $E_{\gamma,peak}^{\prime}=2\ {\rm eV}$ with jet luminosity $L = 10^{51}\ {\rm erg/s}$. We can see that $\gamma_{p,in}$ does not affect the photon output spectra irrespective of the optical depth, which is expected as the timescale at which the electrons are heated due to Coulomb coliisions with protons $t_{e-p}^{\prime} = (\gamma_{e,avg} - 1)m_e c^2/\dot{E}_{e-p}$ is considerably longer than the Comptonization timescale $t_{IC}^{\prime}$. A minimum electron energy $\gamma_{e} \gtrsim 11$ is needed in order to have photons with $E_{\gamma} \gtrsim 10\ {\rm MeV}$ and peak energy $E_{\gamma,peak} \sim 1\ {\rm MeV}$ for both $\tau_{in}$ considered. The output photon spectrum does not show a power-law dependence at both low and high energies when the electron initial energy is small $\gamma_{e,in} \lesssim 11$. While the output photon spectrum shows $\alpha \sim \alpha_{obs}$ and $E_{\gamma,peak} \sim E_{\gamma,obs}$ at both optical depths for electrons with $\gamma_{e,in}=101$ only, the high energy power-law spectral index $|\beta| \gg |\beta_{obs}|$ for $\tau_{in}=40$ and $\sim |\beta_{obs}|$ for $\tau_{in}=20$. It should also be noted that the photon spectrum for $\gamma_{e,in}=3$ looks very similar to the $\gamma_{e,in}=101$ and $N_{rh}=0$ case in Figure \ref{fig1} at both optical depths. This further implies that Coulomb heating of electrons is relatively inefficient and provides insufficient energy to the photons which is analogous to fewer dissipation events occuring in the jet.

\begin{figure*}
  \begin{subfigure}[b]{0.5\linewidth}
    \centering
    \includegraphics[width=0.98\linewidth]{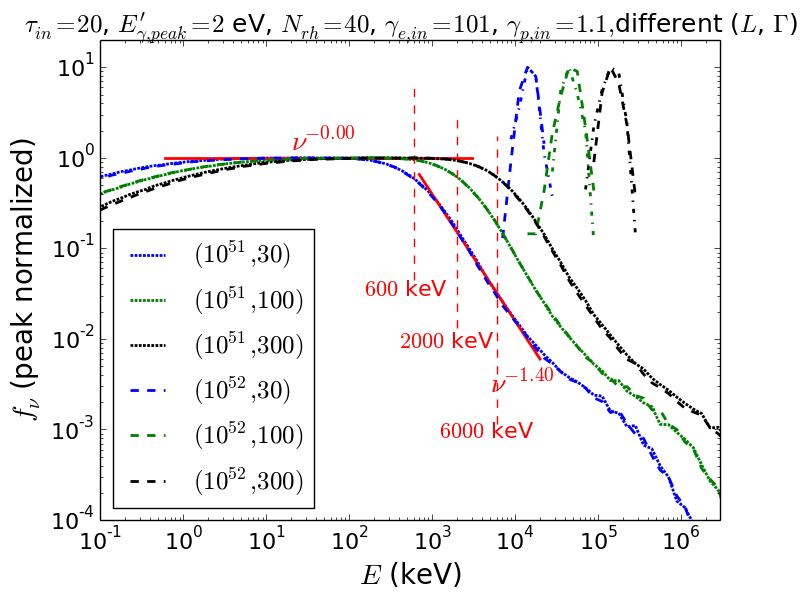}
    %\includegraphics[width=0.97\linewidth]{5.1.ps} 
    %\caption{Initial condition} 
    %\label{fig1:a} 
    %\vspace{1ex}
  \end{subfigure}%%  
  \begin{subfigure}[b]{0.5\linewidth}
   \centering
   \includegraphics[width=0.97\linewidth]{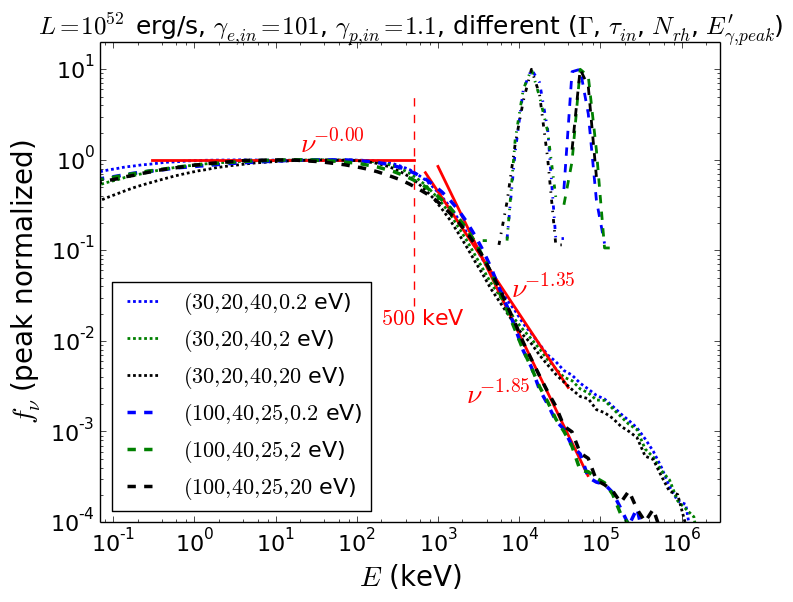} 
   %\includegraphics[width=0.97\linewidth]{5.2.ps}
   %\caption{Initial condition} 
   %\label{fig1:a} 
   %\vspace{1ex}
  \end{subfigure}%% 
  \caption{
  	{\it Left panel:} MCRaT simulations showing the effect of $L=(10^{51},10^{52})\ {\rm erg/s}$ and $\Gamma=30,100,300$ for constant $N_{rh}=40$, $\gamma_{e,in}=101$ and $\tau_{in}=20$. For these simulations, we consider input parameters $E_{\gamma,peak}^{\prime}=2$ eV and $\gamma_{p,in}=1.1$.	 
	{\it Right panel:} MCRaT simulations showing the effect of $E_{\gamma,peak}^{\prime}=0.2,2,20\ {\rm eV}$ for $N_{rh}=40\ (\tau_{in}=20, \Gamma=30)$ and $N_{rh}=25\ (\tau_{in}=40, \Gamma=100)$. For these simulations, we consider input parameters $\gamma_{e,in}=101$, $\gamma_{p,in}=1.1$ and $L=10^{52}\ {\rm erg/s}$. 
	}
  \label{fig4} 
\end{figure*}

In the left panel of Figure \ref{fig4}, we show the simulation results for fixed $E_{inj,cr} = 4000\ m_e c^2$ at $\tau_{in}=20$ and for different combinations of luminosities $L = 10^{51}, 10^{52}\ {\rm erg/s}$ and jet bulk Lorentz factors $\Gamma = 30, 100, 300$. The photons/electrons/protons are initialized with energies $E_{\gamma,peak}^{\prime}=2\ {\rm eV}/\gamma_{e,in}=101/\gamma_{p,in}=1.1$ at optical depth $\tau_{in}=20$. While the jet luminosity $L$ has no noticeable effect on the output photon spectrum, increase in bulk Lorentz factor $\Gamma$ shifts the photon peak energy to higher values. We find that even though $\Gamma$ does not affect $\alpha$ and $\beta$, it rescales photon peak energy as $E_{\gamma,peak} \propto \Gamma$. The output photon spectrum shows $E_{\gamma,peak} \sim E_{\gamma,obs}$ only for smaller $\Gamma \sim 30$ values. While larger $\Gamma \sim 100$ can also reproduce $E_{\gamma,peak} \sim 500\ {\rm keV}$ and $\alpha \sim 0$ at $\tau_{in} \gtrsim 40$ in agreement with the observations, it cannot explain the observed high energy spectral index (see right panel of Figure \ref{fig3}).  

In the right panel of Figure \ref{fig4}, we show the simulation results for fixed $N_{rh,cr}(\tau_{in})=40/25$ at $\tau_{in}=20\ (\Gamma=30)/40\ (\Gamma=100)$ and different seed photon energies $E_{\gamma,peak}^{\prime}=0.2,2,20\ {\rm eV}$. The electrons/protons are initialized with energies $\gamma_{e,in}=101/\gamma_{p,in}=1.1$ with jet luminosity $L=10^{52}\ {\rm erg/s}$. We find that for both $E_{inj,cr}(\tau_{in})=2500\ m_e c^2$ and $4000\ m_e c^2$, the low/high energy spectral index $\alpha/\beta$ and the output photon peak energy $E_{\gamma,peak}$ are practically unaffected by the choice of $E_{\gamma,peak}^{\prime}$. However, there is a noticeable difference in $f_{\nu}$ at very low energies $E_{\gamma} \lesssim \Gamma E_{\gamma,peak}^{\prime}$ for relatively small optical depths $\tau_{in} \lesssim 20$. The specific photon flux $f_{\nu}$ falls off considerably at energies less than $\Gamma E_{\gamma,peak}^{\prime}$ as most of the photons gain energy and do not populate the low energy tail after getting scattered by the electrons. For larger $\tau_{in}$, the photons get scattered multiple times thereby increasing the probability of differential number of scatterings before escaping the photosphere and subsequent broadening of the spectrum. 
As a result, more photons populate the low energy tail of the photon spectrum and the spectra with different initial energies $\Gamma E_{\gamma,peak}^{\prime}$ become indistinguishable for $\tau_{in} \gtrsim 40$. We discuss this geometrical broadening effect in more detail in the next section.

\section{IC spectra for repeated scatterings}
\label{IC_sc}
In the previous section, we obtained the output photon spectrum from MCRaT simulations by including physical processes in the jet such as adiabatic cooling, Coulomb collisions, IC and pair production/annihilation. Here, we will assume that Comptonization is the dominant process influencing the output photon spectrum to first evaluate the energy spectrum of synchrotron photons after they experience single scattering with the electrons. Then we extend this formalism to find the photon energy spectrum for the realistic case when they undergo repeated scatterings with the electrons in the jet before exiting the photosphere. The energy distribution of the scattered photons depends mainly on the incident photon spectrum and the electron energy distribution.

\begin{figure*}  
  \begin{subfigure}[b]{0.5\linewidth}
   \centering
   \includegraphics[width=0.85\linewidth,height=0.65\linewidth]{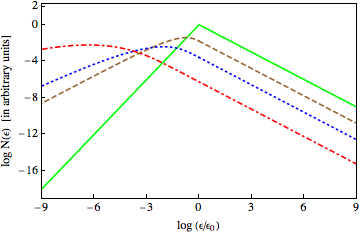} 
   %\includegraphics[width=0.97\linewidth]{5.2.ps}
   %\caption{Initial condition} 
   %\label{fig1:a} 
   %\vspace{1ex}
  \end{subfigure}%% 
  \begin{subfigure}[b]{0.5\linewidth}
   \centering
   \includegraphics[width=0.85\linewidth,height=0.65\linewidth]{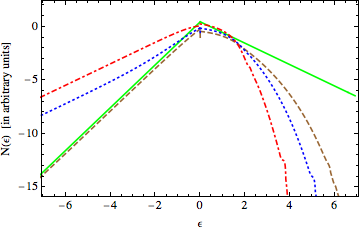} 
   %\includegraphics[width=0.97\linewidth]{5.2.ps}
   %\caption{Initial condition} 
   %\label{fig1:a} 
   %\vspace{1ex}
  \end{subfigure}\\ 
  \begin{subfigure}[b]{0.5\linewidth}
   \centering
   \includegraphics[width=0.97\linewidth]{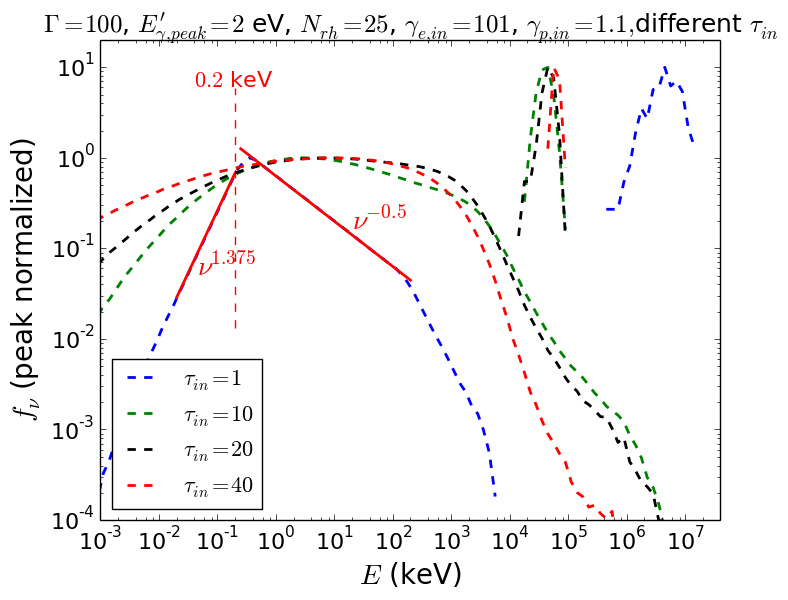} 
   %\includegraphics[width=0.97\linewidth]{5.2.ps}
   %\caption{Initial condition} 
   %\label{fig1:a} 
   %\vspace{1ex}
  \end{subfigure}
  \caption{{\it Effect of geometrical broadening on the photon spectrum for increasing optical depth $\tau_{in}$:}
  {\it Top-left panel:} IC spectrum for fast cooled synchrotron photons with energy $\epsilon_0 = 1$ and mono-energetic electrons with constant energy $\gamma_{e,0} = 1.1$, where the respective energies are in units of $m_e c^2$.
  {\it Top-right panel:} IC spectrum for the same photon seed and Maxwellian electrons with peak energy $\gamma_{e,0} = 1.1$.
  We define $\gamma_{e,0}$ for Maxwellian electrons in terms of the electron temperature $T_{e,0}^{\prime}$ with $k_B T_{e,0}^{\prime} = (\gamma_{e,ad,0}-1)(\gamma_{e,0}-1)m_e c^2$ (see Section \ref{Par_dist}).
  The solid green line, brown dashed line, blue dotted line and red dot-dashed lines are the scattered photon spectra after $N = 0, 1, 2$ and $5$ scatterings, respectively.  
  {\it Bottom panel:} MCRaT simulations showing the effect of geometrical broadening on the photon spectrum at $\tau_{in}=1,10,20,40$ for $(N_{rh},\gamma_{e,in})=(25,101)$ and $\gamma_{p,in}=1.1$. For these simulations, we consider input parameters $L=10^{52}\ {\rm erg/s}$, $E_{\gamma,peak}^{\prime} = 2\ {\rm eV}$ and $\Gamma=100$. 
	}
  \label{fig5} 
\end{figure*}

\subsection{Photon distribution after one scattering}
For this calculation, we will consider electrons and incident photons with isotropic distributions in the jet-comoving frame - in which case the scattered photons are also distributed isotropically in the comoving frame of the jet. For simplicity, we only consider Thomson scattering in the rest frame of the electron and assume that all scattering events are elastic in nature.

For incident photons with energy $\epsilon$ scattering off electrons with energy $\gamma m_e c^2$, the total scattered power per energy per volume is \citep{RL79}
\begin{equation}
\frac{dE}{dVdtd\epsilon_{1}} = \frac{3}{4}c\sigma_{T}\int_{\epsilon_{1}/4\gamma^2}^{\infty} d\epsilon \frac{\epsilon_{1}}{\epsilon^{2}}f(\epsilon)\int_{1}^{\infty} \frac{d\gamma}{\gamma^{2}}n_{e}(\gamma)g_{iso}\left(\frac{\epsilon_{1}}{4\gamma^2 \epsilon}\right),
\label{out_sc}
\end{equation}
where, $\epsilon_{1}$ is the scattered photon energy, $f(\epsilon)$ is the photon distribution function, $n_{e}(\gamma)$ is the electron distribution function and $g_{iso}(x) = \frac{2}{3}(1-x)$ for isotropic photon distribution in the jet-comoving frame.
Here we consider the simple case in which the incident photons have a synchrotron/piecewise power-law energy distribution,
\begin{equation}
f_{in}(\epsilon) = f_{0}\left\{
\begin{array}{ll}
(\epsilon/\epsilon_{0})^{a}, & \epsilon<\epsilon_{0}\\
(\epsilon/\epsilon_{0})^{-b}, & \epsilon>\epsilon_{0}\\
\end{array}
\right. 
\label{FCphseed}
\end{equation}
and the electrons are mono-energetic with $n_{e}(\gamma) = n_{0}\delta(\gamma - \gamma_{0})$. For isotropic photons, equation (\ref{out_sc}) simplifies to
\begin{eqnarray}
\frac{dE}{dVdtd\epsilon_{1}} 
= 2c\sigma_{T}n_{0}\int_{\epsilon_{1}/4\gamma_{0}^{2}}^{\infty}\frac{d\epsilon}{\epsilon}\frac{\epsilon_{1}}{4\gamma_{0}^{2}\epsilon}f(\epsilon)\left(1 - \frac{\epsilon_{1}}{4\gamma_{0}^{2}\epsilon}\right),
\end{eqnarray}
where we have assumed that the electrons are relativistic with $\gamma_0 \gg 1$. Substituting $x=4\gamma_{0}^{2}\epsilon/\epsilon_{1}$ yields
\begin{equation}
\frac{dE}{dVdtd\epsilon_{1}} = 2c\sigma_{T}n_{0}\int_{1}^{\infty}\frac{dx}{x^2}\left(1 - \frac{1}{x}\right)f\left(\frac{\epsilon_{1}x}{4\gamma_{0}^{2}}\right).
\end{equation}

\begin{itemize}

\item {\it For photons below peak energy:} $\epsilon_{1} < 4\gamma_{0}^{2}\epsilon_{0}$ and we can further define $\epsilon_{1}/4\gamma_{0}^{2} = \eta \epsilon_{0}$ with $\eta < 1$ to obtain
\begin{eqnarray}
\frac{dE}{dVdtd\epsilon_{1}} = 2c\sigma_{T}n_{0}f_{0} \nonumber \\
\left[\eta^{a}\int_{1}^{1/\eta}\frac{dx}{x^2}\left(1-\frac{1}{x}\right)x^{a} + \eta^{-b}\int_{1/\eta}^{\infty}\frac{dx}{x^2}\left(1-\frac{1}{x}\right)x^{-b}\right]\nonumber
 \\ 
= 2c\sigma_{T}n_{0}f_{0}\left[\frac{\eta^{a}-\eta}{1-a} - \frac{\eta^{a}-\eta^{2}}{2-a} + \frac{\eta}{b+1} - \frac{\eta^2}{b+2}\right].
\end{eqnarray}
For low energy photons with $\eta \ll 1$, if the incident photons have a hard spectrum with $0 < a < 1$, the scattered photon distribution $f_{sc}(\epsilon_{1}) \propto dE/(dVdtd\epsilon_{1}) \propto \eta^{a} \propto \epsilon_{1}^{a}/(\gamma_{0}^{2a}\epsilon_{0}^{a})$ is the same as that of the incident photons. However, for a softer low energy incident photon spectrum with $a \ge 1$ and $\eta \ll 1$, we obtain $f_{sc}(\epsilon_{1}) \propto dE/(dVdtd\epsilon_{1}) \propto \epsilon_{1}$. Therefore, after single scattering of synchrotron photons with broken power-law energy distribution, the low energy spectrum is unaffected for hard spectra with $a<1$ whereas $f_{sc}(\epsilon) \propto \epsilon$ for softer spectra.

\item {\it For photons above peak energy:} $\epsilon_{1} > 4\gamma_{0}^{2}\epsilon_{0}$ and we define $\eta = \epsilon_1/(4\gamma_0^2 \epsilon_0) > 1$ as earlier to obtain
\begin{eqnarray}
\frac{dE}{dVdtd\epsilon_{1}} = 2c\sigma_{T}n_{0}f_{0}\int_{1}^{\infty}\frac{dx}{x^2}\left(1- \frac{1}{x}\right)(\eta x)^{-b} \propto \eta^{-b} \nonumber \\
\propto \epsilon_{1}^{-b}/(\gamma_{0}^{-2b}\epsilon_{0}^{-b}),
\end{eqnarray}
which is the same as the incident photon spectrum.
\end{itemize} 
For fast cooled synchrotron photon spectrum, we have $a=2$ and $b=-1$, and the scattered photon distribution after single scattering is
\begin{equation}
f_{1}(\epsilon) = f_{sc}(\epsilon) \propto f_{0}\left\{
\begin{array}{ll}
(\epsilon/\epsilon_{0})^{1}, & \epsilon<\epsilon_{0}\\
(\epsilon/\epsilon_{0})^{-1}, & \epsilon>\epsilon_{0}\\
\end{array}
\right. 
\end{equation}
In reality, however, each photon experiences $\sim 2\tau_{in}$ scatterings on an average before escaping the photosphere. Next, we evaluate the photon spectrum for repeated electron-photon scattering events assuming that the electron energy is held constant i.e. for electrons at thermal equilibrium.

\subsection{Photon distribution after repeated scatterings}
With $f_{1}(\epsilon)$ as the incident photon distribution, we can now extend the same formalism to calculate the photon spectrum after subsequent scattering events assuming that the electron and photon distributions remain isotropic in the jet-comoving frame. 

\begin{itemize}

\item {\it After two scatterings per photon} 

The low and the high energy spectrum after each photon in the jet has undergone exactly two scatterings is
\begin{eqnarray}
f_{2,l}(\epsilon_1) = 2c\sigma_{T}n_{0}f_{0}\left(-\eta{\rm ln}\eta + \frac{2}{3}\eta^{2} - \frac{1}{2}\eta\right)
\propto \frac{\epsilon_{1}}{4\gamma_{0}^{2}\epsilon_{0}}{\rm ln}\left(\frac{\epsilon_{1}}{4\gamma_{0}^{2}\epsilon_{0}}\right),\nonumber \\
f_{2,u}(\epsilon_1) = 2c\sigma_{T}n_{0}f_{0}\int_{1}^{\infty}dx\frac{1}{x^2}\left(1 - \frac{1}{x}\right)\eta^{-1}x^{-1} \propto \gamma_{0}^{2}\epsilon_{0}/\epsilon_{1}. \nonumber
\end{eqnarray}
The scattered photon spectrum is then
\begin{equation}
f_{2}(\epsilon) \propto f_{0}\left\{
\begin{array}{ll}
(\epsilon/\epsilon_{0}){\rm ln}(\epsilon/\epsilon_{0}), & \epsilon<\epsilon_{0}\\
(\epsilon/\epsilon_{0})^{-1}, & \epsilon>\epsilon_{0}\\
\end{array}
\right. 
\end{equation}

\item {\it After three scatterings per photon}

After each photon has undergone exactly three scatterings, the low and high energy are given as
\begin{eqnarray}
f_{3,l}(\epsilon_1) = 2c\sigma_{T}n_{0}f_{0}\nonumber \\
\left[\frac{1}{2}\eta({\rm ln}\eta)^{2} - \eta(\eta+{\rm ln}\eta){\rm ln}\eta + \left(\frac{2}{3}+{\rm ln}\eta\right)\eta^{2} - \left({\rm ln}\eta + \frac{1}{2}\right)\eta\right] \nonumber \\ 
\propto \frac{\epsilon_{1}}{4\gamma_{0}^{2}\epsilon_{0}}\left[{\rm ln}\left(\frac{\epsilon_{1}}{4\gamma_{0}^{2}\epsilon_{0}}\right)\right]^{2}, \nonumber \\
f_{3,u}(\epsilon_1) = 2c\sigma_{T}n_{0}f_{0}\int_{1}^{\infty}dx\frac{1}{x^2}\left(1 - \frac{1}{x}\right)\eta^{-1}x^{-1}
\propto \gamma_{0}^{2}\epsilon_{0}/\epsilon_{1}, \nonumber
\end{eqnarray}
and the scattered photon spectrum is
\begin{equation}
f_{3}(\epsilon) \propto f_{0}\left\{
\begin{array}{ll}
(\epsilon/\epsilon_{0})\left[{\rm ln}(\epsilon/\epsilon_{0})\right]^{2}, & \epsilon<\epsilon_{0}\\
(\epsilon/\epsilon_{0})^{-1}, & \epsilon>\epsilon_{0}\\
\end{array}
\right. 
\end{equation}

\item {\it After $N$ scatterings per photon}

Using similar algebra, it can be shown that after four scatterings per photon, $f_{4,l}(\epsilon) \propto (\epsilon/\epsilon_0)[{\rm ln}(\epsilon/\epsilon_0)]^{3}$ and $f_{4,u}(\epsilon) \propto (\epsilon/\epsilon_0)^{-1}$. We can generalize the above results further for $N \sim 2\tau_{in}$ scatterings per photon and write, 
\begin{equation}
f_{N}(\epsilon) \propto f_{0}\left\{
\begin{array}{ll}
(\epsilon/4\gamma_{0}^{2}\epsilon_{0})\left[{\rm ln}(\epsilon/4\gamma_{0}^{2}\epsilon_{0})\right]^{N-1}, & \epsilon<\epsilon_{0}\\
(\epsilon/4\gamma_{0}^{2}\epsilon_{0})^{-1}, & \epsilon>\epsilon_{0}\\
\end{array}
\right. 
\label{sc_N}
\end{equation}
\end{itemize}

In Figure \ref{fig5}, we show how the photon spectrum is affected by Comptonization with electrons as $\tau_{in}$ and number of scatterings increase. In the top-left/right panel, the IC scattered photon spectrum for fast cooled synchrotron seed photons (Equation \ref{FCphseed}, with $a=2$ and $b=-1$) with energy $\epsilon_0 = m_e c^2$ and mono-energetic/Maxwellian electrons with peak energy $\gamma_{e,0}=1.1$ are shown for scattering orders $N=0,1,2,5$. 
The peak energy $\gamma_{e,0}$ for Maxwellian electrons is defined in terms of the electron temperature $T_{e,0}^{\prime}$ in the jet-comoving frame as $k_B T_{e,0}^{\prime} = (\gamma_{e,ad,0}-1)(\gamma_{e,0}-1)m_e c^2$.
As predicted by equation (\ref{sc_N}), the photon spectrum becomes gradually softer below peak energy as the scattering order increases for both cases. While the high-energy spectrum is power-law $f_{\nu} \propto \nu^{-1}$ irrespective of $N$ for mono-energetic electrons, $f_{\nu} \propto e^{-\nu}$ at high energies for Maxwellian electrons for larger $N$ . This difference is expected as photons scattering off Maxwellian electrons with fixed energy get thermalized at equilibrium to attain a high-energy exponential tail for large optical depths/scatterings. It should be noted that even though the nature of the photon spectrum differs at high energies for these two cases, the qualitative effect is very similar at low energies - gradual flattening of the low-energy spectrum with increase in scattering order $N$. This physical behaviour as predicted by equation (\ref{FCphseed}) can robustly explain the low-energy non-thermal behaviour of the observed photon spectrum, even without other physical processes such as adiabatic cooling, Coulomb collisions and energy injection through dissipation events.

In the bottom panel of Figure \ref{fig5}, we present the MCRaT simulation results for $E_{inj,cr} = 2500\ m_e c^2$ and different optical depths $\tau_{in}=1,10,20,40$. The number of repeated dissipation events in the jet are $N_{rh}=25$ with initial photon/electron/proton energy $E_{\gamma,peak}^{\prime}=2\ {\rm eV}$/$\gamma_{e,in} = 101$/$\gamma_{p,in} = 1.1$ and jet parameters $L=10^{52}\ {\rm erg/s}$ and $\Gamma=100$. It should be noted that unlike the scattered photon spectra obtained from the analytical expression in equation (\ref{out_sc}) in the top two panels of Figure \ref{fig5}, the MCRaT simulation results in the bottom panel include both electron heating (Coulomb interaction and dissipation events) and adiabatic cooling effects. With increase in scattering order ($\propto \tau_{in}$), the high energy photon spectrum becomes steeper with a simultaneous decrease in $E_{\gamma,peak}$. These photons then populate the low energy spectrum and extend the non-thermal tail to energies much lower than $E_{\gamma,peak} \sim 0.2\ {\rm keV}$. The photon spectra from simulations are also considerably broader compared to the analytical results for similar values of $N$. This is directly related to the fact that the photon spectra obtained from MCRaT simulations are nothing but the averaged scattered photon spectrum
\begin{eqnarray}
f_{avg}(\epsilon) \propto \sum_{K=0}^{N_{sc,max}}P(N_{sc}=K)f_{K}(\epsilon),
\label{f_avg}
\end{eqnarray}
where, $f_{K}(\epsilon)$ given by equation (\ref{sc_N}) is the scattered photon spectrum after exactly $K$ scatterings for each photon and $P(N_{sc}=K)$ is the probability for a photon to get scattered exactly $K$ times which is given by \citep{Poz83,ST80},
\begin{eqnarray}
P(N_{sc}=K) \propto \left\{
\begin{array}{ll}
\frac{N_{sc,avg}}{K^{3/2}}\ {\rm exp}\left(-\frac{3N_{sc,avg}}{4K}\right), \ K < N_{sc,avg}\\
\frac{1}{N_{sc,avg}}\ {\rm exp}\left[-\frac{K\pi^2}{3N_{sc,avg}}\right], \ K > N_{sc,avg}
\end{array}
\right. 
\end{eqnarray}
Here, $N_{sc,avg} \sim 2\tau_{in}$ is the average number of scatterings per photon at an optical depth $\tau_{in}$. While the probability of a particular photon getting scattered much larger or much smaller number of times compared to $N_{sc,avg}$ reduces exponentially, there can still be considerable contribution from different scattering orders leading to significant broadening of the photon spectrum.

\subsection{Photon spectrum due to unsaturated Comptonization}
Here we consider the situation when Comptonization is important but the photon spectrum does not saturate to the equilibrium Wien distribution for the majority of the photons in the jet as the electrons cannot supply sufficient energy due to their small $T_{e}^{\prime}$. 
In the absence of a photon source other than the fast-cooled electrons accelerated close to the central engine, the time evolution of the isotropic photon phase space density $n(\epsilon)$ due to scattering from electrons can be estimated with the Boltzmann equation \citep{RL79}
%modified, steady-state Kompaneet's equation 
\begin{eqnarray}
\frac{1}{c}\frac{\partial{n(\epsilon)}}{\partial t} = \int d^3 p \int d\Omega \frac{d\sigma}{d\Omega} [f_e({\bf p_1})n(\epsilon_1)(1 + n(\epsilon)) \nonumber \\ 
- f_e({\bf p})n(\epsilon)(1 + n(\epsilon_1))],
\label{Boltz}
\end{eqnarray}
%\begin{eqnarray}
%\frac{1}{x^2} \frac{\partial}{\partial x} \left[x^4 \left(\frac{\partial n}{\partial x} + n\right) \right] = \frac{4n}{y},
%\label{Komp}
%\end{eqnarray}
where $d\sigma/d\Omega$ is the scattering cross-section, ${\bf p}$/${\bf p_1}$ is the incident/scattered electron momentum, $\epsilon$/$\epsilon_1$ is the incident/scattered photon energy and $f_e(p)$ is the phase space density of non-relativistic thermal electrons.
%where $x = \epsilon/kT_{e}^{\prime}$ and the Compton-Y parameter $y = (4kT_{e}^{\prime}/m_e c^2){\rm max}(\tau_{in},\tau_{in}^{2})$. 
As the fractional energy transfer per scattering is considerably small with $\Delta = (\epsilon_1 - \epsilon)/kT_e^{\prime} \ll 1$ for non-relativistic electrons, the Boltzmann equation can be expanded to second order in $\Delta$ using
\begin{eqnarray}
n(\epsilon_1) \approx n(\epsilon) + kT_e^{\prime} \Delta \frac{\partial n}{\partial \epsilon} + \frac{1}{2}(kT_e^{\prime} \Delta)^2 \frac{\partial^2 n}{\partial \epsilon^2}, \nonumber \\
f_e(E_1) \approx f_e(E) + kT_e^{\prime} \Delta \frac{\partial f_e}{\partial E} + \frac{1}{2}(kT_e^{\prime} \Delta)^2 \frac{\partial^2 f_e}{\partial E^2}, \nonumber
\label{Taylor_exp}
\end{eqnarray}
where $E = p^2/2m_e$ is the electron energy. Substituting the Taylor expansions of $n(\epsilon_1)$ and $f_e(E_1)$ into the Boltzmann equation, and further assuming elastic scattering simplifies equation (\ref{Boltz}) to 
\begin{eqnarray}
\frac{4n}{y} = \left(\frac{\epsilon}{kT_e^{\prime}}\right)^{2} \left[(kT_e^{\prime})^{2} \frac{\partial^2 n}{\partial \epsilon^2} + kT_e^{\prime} \frac{\partial n}{\partial \epsilon} \right] \nonumber \\
+\ 4 \left(\frac{\epsilon}{kT_e^{\prime}}\right) \left[kT_e^{\prime} \frac{\partial n}{\partial \epsilon} + n \right].
\label{modif_Komp}
\end{eqnarray}
Here we have ignored the stimulated emission term and used $y = (4kT_{e}^{\prime}/m_e c^2)\tau_{in}$ as the Compton-Y parameter. 

For very large photon energies $\epsilon/kT_e^{\prime} \gg 1$, the photon spectrum falls off exponentially with $n(\epsilon) \propto {\rm exp}(-\epsilon/kT_e^{\prime})$ being an approximate solution to equation (\ref{modif_Komp}). 
%In equation (\ref{Komp}), we assume that the electrons cool down to non-relativistic energies $\gamma_{e} \sim \gamma_{e,Comp}$ rapidly and ignore the $n^2$ term from stimulated emission. 
%For very large photon energies corresponding to $x \gg 1$, the photon spectrum falls off exponentially with $n(x) \propto {\rm exp}(-x)$ being an approximate solution to equation (\ref{Komp}). 
However, $y \gtrsim 1$ can still be sufficient in order to populate the power-law photon spectrum $n(\epsilon) \propto \epsilon^{-1}$ just above the peak energy $\epsilon_{peak}$ before the electrons rapidly cool down to non-relativistic energies $\sim \gamma_{e,Comp}$ \citep{Santana16}. 
For considerably smaller photon energies $\epsilon/kT_e^{\prime} \ll 1$, the recoil term $n$ can be neglected in comparison to the upscattering term $kT_e^{\prime} \partial n/\partial x$ and the general solution is then power-law $n(\epsilon) \propto (\epsilon/kT_e^{\prime})^{p}$ with 
\begin{eqnarray}
p = -\left(1.5 + \frac{\epsilon}{2kT_e^{\prime}}\right) \pm \sqrt{\left(1.5 + \frac{\epsilon}{2kT_e^{\prime}}\right)^{2} + \frac{m_e c^2}{k T_e^{\prime}}\frac{1}{\tau_{in}}}. \nonumber
\end{eqnarray} 
While the larger (smaller) root is appropriate for $y \gg 1$ ($y \ll 1$), a linear combination of both is valid for $y \sim 1$. In the presence of dissipation events occurring in the jet, $T_{e}^{\prime}$ is elevated by a factor $\xi = [1 + (t_{dyn}^{\prime}/t_{diss}^{\prime}) - (t_{dyn}^{\prime}/t_{IC}^{\prime})]^{\alpha}$ for $\alpha \geq 0$, with $t_{diss}^{\prime} \approx t_{dyn}^{\prime}/N_{rh}$ being the characteristic energy dissipation timescale. As expected, $T_{e}^{\prime}$ increases (decreases) with a reduction in $t_{diss}^{\prime}$ ($t_{IC}^{\prime}$) and is unaffected by dissipation for $t_{diss}^{\prime} \approx t_{IC}^{\prime}$. 
To obtain $f_{\nu} \propto \nu^0$ for photon energies below $\epsilon_{peak}$, we need to have $n(\epsilon) \propto (\epsilon/kT_{e}^{\prime})^{-3}$. In order to satisfy this criterion, we require 
\begin{eqnarray}
\frac{kT_{e}^{\prime}}{m_e c^2} \left[1 + \left(\frac{t_{dyn}^{\prime}}{t_{diss}^{\prime}} - \frac{t_{dyn}^{\prime}}{t_{IC}^{\prime}}\right)\right]^{\alpha} \tau_{in} \gg 1 \nonumber
\end{eqnarray}
As $kT_{e}^{\prime}/m_e c^2 \sim 1$ for non-relativistic electrons and $\tau_{in} \sim {\rm few}-10$, this implies $(t_{dyn}^{\prime}/t_{diss}^{\prime} - t_{dyn}^{\prime}/t_{IC}^{\prime}) \gg 1$ that is
\begin{eqnarray}
N_{rh} \left(1 - \frac{t_{dyn}^{\prime}}{N_{rh}t_{IC}^{\prime}}\right) \gg 1. \nonumber
\end{eqnarray}
Therefore, a flat non-thermal photon spectrum can be obtained at low energies for sufficiently large $N_{rh} \gtrsim 10$ as $t_{dyn}^{\prime} \approx t_{IC}^{\prime}$ once the electrons cool down to energies $\sim \gamma_{e,Comp}$. From the evolution of $\gamma_{e}$ with scattering order as shown in the right-half panels of Figure \ref{fig1}, we know that the electrons rapidly cool down to sub-relativistic energies even in the presence of repeated energy dissipation events. 
%For considerably smaller photon energies with $x \ll 1$, the recoil term $n$ can be neglected in comparison to the upscattering term $\partial n/\partial x$ in the Kompaneet's equation and the general solution is then power-law $n(x) \propto x^{p}$ with $p = -1.5 \pm \sqrt{2.25 + (4/y)}$. 
%The general solution to equation (\ref{Komp}) is still a power-law distribution with an index $q = -1.5 \pm \sqrt{2.25 + (4/\xi y)}$. 
% $n(x) \propto x^{-3}$ corresponding to $\xi y \gg 1$ for $y \sim 1$ is needed. 
%This condition is always satisfied when 

\section{Discussions and Conclusions}
\label{disc_conc}
In this paper, we explored the photospheric emission model in detail to better understand the GRB prompt emission radiation mechanism. The primary objective was to utilize our MCRaT photospheric code to explain the distinct non-thermal behaviour of the prompt emission spectrum, $f_{\nu} \propto \nu^{0}/f_{\nu} \propto \nu^{-1.2}$ at low/high photon energies along with observed peak energy at $E_{\gamma,peak} \sim 300\ {\rm MeV}$. For all our simulations, we have considered Comptonization of fast cooled synchrotron photons with Maxwellian electrons and for photon to electron number ratio $N_{\gamma}/N_{e} \sim 10^5$, consistent with observed radiation efficiency. The electrons in the jet are accelerated and maintained at certain critical energy by two different mechanisms: 1. continuous energy transfer via Coulomb collisions with mono-energetic seed protons, 2. repeated episodic energy dissipation events that are equally spaced over scatterings and accelerate electrons and protons back to their initial energies. 

\begin{figure}
  \begin{subfigure}[b]{\linewidth}
    \centering
    \includegraphics[width=0.97\linewidth]{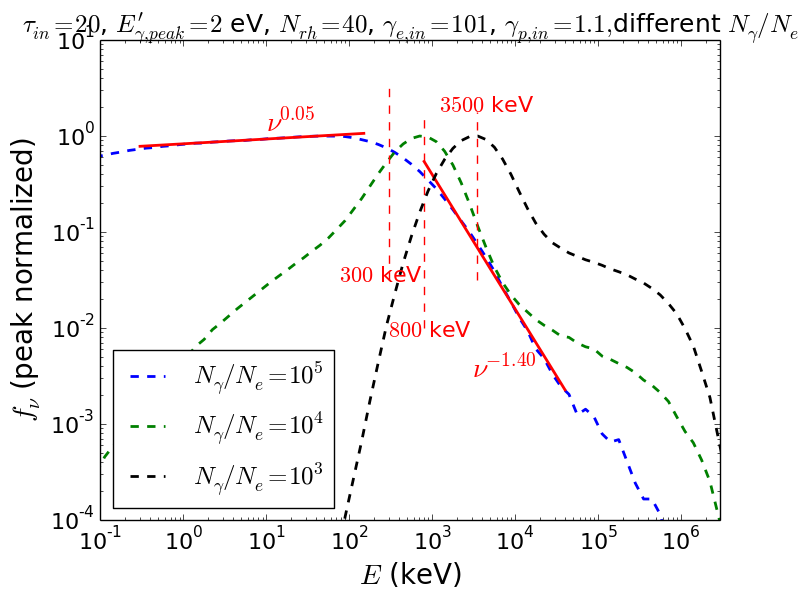}
    %\includegraphics[width=0.97\linewidth]{5.1.ps} 
    %\caption{Initial condition} 
    %\label{fig1:a} 
    %\vspace{1ex}
  \end{subfigure}%%  
  \caption{MCRaT simulations showing the effect of $N_{\gamma}/N_{e} = 10^7/10^2,10^7/10^3,10^7/10^4$ for constant $N_{rh}=40$, $\gamma_{e,in}=101$ and $\tau_{in}=20$. For these simulations, we consider input parameters $\gamma_{p,in}=1.1$, $L=10^{52}\ {\rm erg/s}$, $E_{\gamma,peak}^{\prime}=2\ {\rm eV}$ and $\Gamma=30$.   
	}
  \label{fig6} 
\end{figure}

\begin{figure}
  \begin{subfigure}[b]{\linewidth}
    \centering
    \includegraphics[width=0.93\linewidth]{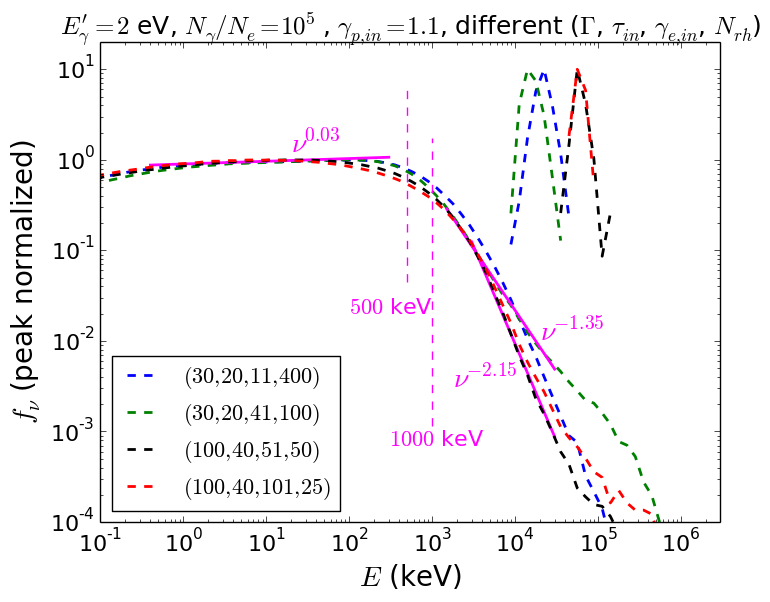}
    %\includegraphics[width=0.97\linewidth]{5.1.ps} 
    %\caption{Initial condition} 
    %\label{fig1:a} 
    %\vspace{1ex}
  \end{subfigure}%%  
  \caption{{\it MCRaT simulation results with the best set of parameters for a jet with $N_{\gamma}/N_{e}=10^5$.}
The relativistic jet with $L=10^{52}\ {\rm erg/s}$ has photons with $E_{\gamma,peak}^{\prime}=2\ {\rm eV}$ and protons with $\gamma_{p,in}=1.1$. The energy injection necessary in order to produce an output photon spectrum with the observed Band-like spectral properties depends on $\tau_{in}$ and $\Gamma$. Here we consider $E_{inj}=4000/2500\ m_e c^2$ for $\Gamma=30/100$ and $\tau_{in}=20/40$, for two distinct electron energies $\gamma_{e,in}=(11,41)/(51,101)$.  
  } 
  \label{fig7} 
\end{figure}

In order to scatter synchrotron seed photons with energy $\Gamma E_{\gamma,peak}^{\prime} \lesssim 1\ {\rm keV}$ to energies $E_{\gamma,obs} \gtrsim 300\ {\rm keV}$ and populate the high energy power-law tail with $f_{\nu} \propto \nu^{-1.2}$, the electron kinetic energy during jet expansion should at least be larger than the energy requirement of the photons. While the initial kinetic energy of the electrons is $(\gamma_{e,in}-1)m_e c^2$, the protons transfer part of their kinetic energy $\sim (t_{dyn}^{\prime}/t_{Coul}^{\prime})(\gamma_{p,in}-1)m_p c^2$ to the electrons and the sub-photospheric dissipation events inject an additional energy $E_{inj} \sim N_{rh}(\gamma_{e,in}-1)m_e c^2$ into the electrons until the outflow becomes so optically thin that the photons can escape through the photosphere. As the photons experience roughly $\sim \tau_{in}t_{dyn}^{\prime}/t_{IC}^{\prime}$ scatterings before escaping and the jet is charge neutral ($N_{e}=N_{p}$),
\begin{eqnarray}
\frac{N_{\gamma}}{N_{e}}E_{\gamma,avg}^{\prime} \approx \nonumber \\
\left[\frac{t_{dyn}^{\prime}}{t_{Coul}^{\prime}}(\gamma_{p,in}-1)m_p c^2 + N_{rh}(\gamma_{e,in}-1)m_e c^2\right]\frac{\tau_{in}t_{dyn}^{\prime}}{t_{IC}^{\prime}},
\label{Nph_Ne}
\end{eqnarray}
where we assume that the timescales are roughly constant once the electrons and protons attain their equilibrium energies. In Figure \ref{fig6}, we present the simulation results for $N_{rh,cr}=40$ at $\tau_{in}=20$ for different photon to electron number ratios $N_{\gamma}/N_{e} = 10^7/10^4, 10^7/10^3, 10^7/10^2$. The photons/electrons/protons are initialized with energies $E_{\gamma,peak}^{\prime}=2\ {\rm eV}$/$\gamma_{e,in}=101$/$\gamma_{p,in}=1.1$ for jet parameters $L = 10^{52}\ {\rm erg/s}$ and $\Gamma=30$. We find that $E_{\gamma,peak}$ shifts to larger energies $\gtrsim 1\ {\rm MeV}$ and photons have more energy on average as the number ratio $N_{\gamma}/N_{e}$ decreases. This is expected from equation (\ref{Nph_Ne}) as more electrons for a given photon number means larger energy injection into the photons for similar jet parameters. Moreover, it is easier to scatter photons to very large energies and extend the power-law tail $f_{\nu} \propto \nu^{-1.2}$ to few $100\ {\rm MeV}$ energies even without episodic energy injection events in the jet \citep{Santana16,MB18}. As $E_{inj,tot} \propto N_{rh,crit}N_{e}(\gamma_{e,in}-1)$, the observed low energy spectral index $\alpha_{obs} \sim 0$ may also be achieved with either smaller $N_{rh,crit}$ or smaller $\gamma_{e,in}$ for smaller number ratios $N_{\gamma}/N_{e}$ and with considerable geometrical broadening for large $\tau_{in}$. However, previous MCRaT photospheric simulations with relatively smaller $N_{\gamma}/N_{e} \sim 10^1 - 10^4$ could not successfully explain the flat low energy photon spectrum \citep{LB10,CL15}. 

In Section \ref{sim_results}, we studied in detail the effect of jet parameters and particle energies on the output photon spectrum. The parameters that significantly affect the spectral properties for a given $N_{\gamma}/N_{e}$ are $E_{inj}(\gamma_{e,in},N_{rh})$, $\Gamma$ and $\tau_{in}$. In Figure \ref{fig7}, we present the simulation results for the most probable set of parameters that gives output photon spectrum with $(\alpha,\beta,E_{\gamma,peak})$ very similar to the observed GRB prompt emission spectrum. The photons/protons in these simulations are initialized with energies $E_{\gamma,peak}^{\prime}=2\ {\rm eV}$/$\gamma_{p,in}=1.1$ for jet parameters $L = 10^{52}\ {\rm erg/s}$, $N_{\gamma}/N_{e}=10^5$ and $\Gamma \sim 30-100$. The particles are injected with energy $E_{inj,cr} \sim 2500-4000\ m_e c^2$ per electron for a range of optical depth $\tau_{in} \sim 20-40$. 
For smaller optical depths $\tau_{in} \sim 20$ and jet bulk Lorentz factor $\Gamma \sim 30$, $(\alpha,\beta,E_{\gamma,peak}) \sim (0,-1.4,1\ {\rm MeV})$ is obtained with $E_{inj,cr} \sim 4000\ m_e c^2$ and $\gamma_{e,in} \gtrsim 40$. Although $\alpha \sim 0$ and $E_{\gamma,peak} \sim 500\ {\rm keV}$ for $E_{inj,cr} \sim 2500\ m_e c^2$ at larger $\tau_{in} \sim 40$ and $\Gamma \sim 100$, the high energy spectrum is significantly steeper than the observed prompt spectrum with $\beta \sim -2.1$, especially for $\gamma_{e,in} \lesssim 50$. For a fixed $E_{inj,cr}(\tau_{in})$, while $\beta$ spans a broader range with variation in $\gamma_{e,in} \sim 10-40$ at smaller $\tau_{in} \sim 20$, it is relatively independent of $\gamma_{e,in}$ for larger $\tau_{in} \sim 40$. This is in perfect agreement with the theoretical predictions of the photospheric emission model as the shape of the output photon spectrum is almost entirely determined by the number of scatterings ($\propto \tau_{in}$) with the initial particle energies becoming progressively unimportant in the limit of large optical depths.

Here we summarize the main results of this work:
%\begin{enumerate}
%\item 

(i) The electrons cool down very rapidly to non-relativistic energies ($N_{Comp} \sim 10^{4}$, $t \sim 10^{-2}\ t_{dyn}$) in the absence of any external dissipation events. As $t_{IC} \lesssim t_{Coul} \ll t_{dyn}$, the electrons attain equilibrium with energy $\gamma_{e} \sim \gamma_{e,eq} \ll \gamma_{e,crit}$ after $\sim N_{Comp}$ scatterings and cannot scatter the bulk of the photons to $\sim {\rm MeV}$ energies. This entails energy injection into the jet particles via either (continuous) super-efficient Coulomb collisions or (episodic) sub-photospheric dissipation events. However, for the Coulomb heating efficiencies necessary, the protons lose a considerable fraction of their energy within jet expansion timescales $\sim t_{dyn}$, for $\tau_{in} \gtrsim 10$, to attain non-relativistic energies comparable to that of the electrons. As a result, continuous energy injection by protons is not sufficient to maintain electrons at $\gamma_{e} \sim \gamma_{e,crit}$ and produce the observed photon spectrum, especially for larger optical depths.

(ii) The required energy injection can rather be achieved with episodic sub-photospheric dissipation events through a variety of mechanisms such as internal shocks, magnetic reconnections, neutron-proton collisions, etc. These events can keep the electrons at energies $\gamma_{e} \gtrsim \gamma_{e,crit}$ provided that they are sufficiently energetic and frequent. We find that a $E_{inj}-\tau_{in}$ correlation is essential for the electrons to scatter the jet photons to observed energies $E_{\gamma,obs}$: for large $E_{inj}$, the photon peak energy $E_{\gamma,peak} \gg E_{\gamma,obs} \sim 300\ {\rm keV}$, while for large $\tau_{in}$, $E_{\gamma,peak} \ll E_{\gamma,obs}$ due to significant adiabatic loss. While this is a necessary condition to determine the average photon energy in the observed spectrum, it is not sufficient to constrain its general non-thermal shape. From MCRaT simulations, we quantify the $E_{inj} - \tau_{in}$ correlation: $E_{inj,cr} = 6000/4000/2500\ m_e c^2$ per electron for $\tau_{in} = 10/20/40$ to determine the effect of energy injection on the Comptonized output photon spectra.

(iii) In the output photon spectrum, $\alpha$ critically depends on $E_{inj}$ whereas $\beta$ and $E_{\gamma,peak}$ are almost entirely determined by $\tau_{in}$ (independent of $N_{rh}$). With an increase in $\tau_{in}$, $E_{\gamma,peak}$ decreases and the high-energy photon spectrum becomes steeper. Additionally, $|\beta|$ also increases with decrease in initial electron energy $\gamma_{e,in}$ for fixed $E_{inj,cr} = N_{rh,cr}(\gamma_{e,in}-1)\ m_e c^2$. In order to have $E_{\gamma,peak} \sim E_{\gamma,obs}$ and $|\beta| \sim |\beta|_{obs}$, particles and photons need to be initialized at $\tau_{in} \sim 20-40$ and injected with energy $E_{inj,cr} \sim 2500-4000\ m_e c^2$ for $\gamma_{e,in} \sim {\rm few}\ 10{\rm s}$. Initial proton energy $\gamma_{p,in}$ does not influence photon spectrum irrespective of $\tau_{in}$ - which is expected as electron heating timescale $\gg t_{IC}$. The jet luminosity $L$ has no appreciable effect on the photon spectrum whereas photon peak energy scales directly with the jet bulk Lorentz factor, $E_{\gamma,peak} \propto \Gamma$. We find that $E_{\gamma,peak} \sim E_{\gamma,obs}$ only for smaller $\Gamma \sim 30$ - while larger $\Gamma \sim 100$ gives $E_{\gamma,peak} \sim 500\ {\rm keV}$ at $\tau_{in} \sim 40$, the high energy photon spectrum is considerably steeper than observed. The seed photon energy $E_{\gamma,peak}^{seed}$ is relatively unimportant and only affects observed photon flux $f_{\nu}$ for very low energies at smaller $\tau_{in} \lesssim 20$.

(iv) For isotropic electrons scattering isotropic photons, the scattered photon energy distribution is isotropic and can be analytically evaluated for lower order scatterings and for a given electron and photon energy distribution. We show that a non-thermal photon spectrum with $\alpha \sim 0$ and $\beta \sim -1$ is obtained for mono-energetic electrons scattering fast cooled synchrotron photons at moderate optical depths. For Comptonization of synchrotron photons with Maxwellian electrons, $\alpha \sim 0$ behaviour is retained at low energies whereas $f_{\nu} \propto e^{-\nu}$ at high energies. The output photon spectrum is essentially the scattered photon spectra averaged with the relevant scattering probability distribution. Qualitatively, the low-energy non-thermal dependence $\alpha \sim 0$ is obtained from multiple scatterings and subsequent geometrical broadening of the spectrum whereas the high-energy power-law dependence is primarily attributed to repeated episodic and continuous energy injection events in the relativistic jet.

(v) The spectral parameters $(\alpha,\beta,E_{\gamma,peak})$ of the observed GRB prompt emission spectrum can be robustly explained with: sub-photospheric Comptonization of fast cooled synchrotron photons while electrons and protons are accelerated to relativistic energies due to repeated dissipation events. Sub-relativistic protons continuously heat up the electrons via Coulomb collisions in a relativistic jet with $\Gamma \sim 30$, $L \sim 10^{52}\ {\rm erg/s}$ and $N_{\gamma}/N_{e} \sim 10^5$. The seed synchrotron photons/Maxwellian electrons/mono-energetic protons are injected at moderate optical depths $\tau_{in} \sim 20$ with energies $E_{\gamma,peak}^{\prime} \sim 2\ {\rm eV}$/$\gamma_{e,in} \sim 50$/$\gamma_{p,in} \sim 1.1$. The jet particles are episodically accelerated by dissipation events that are equally spaced over scatterings and inject energy $E_{inj,cr} \sim 4000\ m_e c^2$. We find that both low and high-energy non-thermal observed spectra $(\alpha,\beta,E_{\gamma,peak}) \sim (0,-1.4,1\ {\rm MeV})$ are obtained for smaller optical depths $\tau_{in} \sim 20$ and $\Gamma \sim 30$ when electrons with energy $\gamma_{e,in} \gtrsim 40$ are injected with $E_{inj,cr} \sim 4000\ m_e c^2$. However, for larger $\tau_{in} \sim 40$ and $\Gamma \sim 100$, even though $\alpha \sim 0$ and $E_{\gamma,peak} \sim 500\ {\rm keV}$, the high-energy spectrum is considerably steeper with $|\beta| \sim 2.1 > |\beta|_{obs}$ for $\gamma_{e,in} \lesssim 50$ and $E_{inj,cr} \sim 2500\ m_e c^2$.

%\vspace{0.1 in}

\section*{Acknowledgments}
%\begin{comment}
We thank Paz Beniamini and Bing Zhang for useful discussions. MB would like to thank Milos Milosavljevic for generously providing the computational facilities required for this work.

%\end{comment}

\begin{appendix}

\section{Pair production and annihilation algorithm}
In this Appendix, we describe the algorithm that we implement for pair production and annihilation processes in the jet. All random numbers are drawn from the uniform distribution in the interval 0 to 1. Bold-faced characters denote vectors and $\hat{x}$, $\hat{y}$, $\hat{z}$ are the unit vectors in Cartesian coordinates.
 
 \subsection{Pair production}
The photons are stored in a priority queue ordered by increasing values of travel distances and before every scattering event the photon at the top of this queue with energy $E_{\gamma,1}^{\prime}$ and direction ${\bf \Omega_{1}^{\prime}}=(\Omega_{1,1}^{\prime},\Omega_{2,1}^{\prime},\Omega_{3,1}^{\prime})$ is propagated. The random direction ${\bf \Omega^{\prime}}$ for a photon is initialized using the algorithm described in Appendix C1 of \citet{Santana16}. For pair production, after each scattering event the energy $E_{\gamma,2}^{\prime}$ and direction ${\bf \Omega_{2}^{\prime}}=(\Omega_{1,2}^{\prime},\Omega_{2,2}^{\prime},\Omega_{3,2}^{\prime})$ of the second photon in the priority queue is also extracted. The pair production cross section is \citep{Poz83},
 \begin{eqnarray}
 \sigma_{\gamma\gamma} = \frac{3}{8}\frac{\sigma_{T}}{y^2}\nonumber \\
 \left[\left(2+\frac{2}{y^2}-\frac{1}{y^4}\right){\rm ln}(y+\sqrt{y^2 -1}) - \left(1+\frac{1}{y^2}\right)\left(1-\frac{1}{y^2}\right)^{1/2}\right],
 \label{pp_cs}
 \end{eqnarray}
where, $y^{2}=0.5(E_{\gamma,1}^{\prime}/m_{e}c^2)(E_{\gamma,2}^{\prime}/m_{e}c^2)(1-\rm{cos\ }\theta)$ is a dimensionless energy parameter and $\theta = \Omega_{1,1}^{\prime}\Omega_{1,2}^{\prime} + \Omega_{2,1}^{\prime}\Omega_{2,2}^{\prime} + \Omega_{3,1}^{\prime}\Omega_{3,2}^{\prime}$ is the angle between the incoming photons.

To determine whether pair production event will occur, we draw a random number $\xi_{p}$. Pair production from the selected photons takes place only if $\xi_{p} \leq \sigma_{\gamma \gamma}/\sigma_{T}$ is satisfied. After every such event, an electron and a positron are generated and the photons are not pushed back to the priority queue. Next we draw a random number $\xi_{pE}$ to calculate the energies of the outgoing electron and positron with the expressions: $\gamma_{e} = (\xi_{pE}/m_e c^2)(E_{\gamma,1}^{\prime} + E_{\gamma,2}^{\prime})$ and $\gamma_{pos} = ((1-\xi_{pE})/m_e c^2)(E_{\gamma,1}^{\prime} + E_{\gamma,2}^{\prime})$. The direction of the outgoing electron is evaluated from random numbers $\xi_{1v}$ and $\xi_{2v}$ as
\begin{eqnarray}
%\begin{align}
&v_{3,e}^{\prime} = 2\xi_{1v} - 1, \nonumber \\
&v_{2,e}^{\prime} = \sqrt{1 - v_{3,e}^{\prime 2}}\ {\rm sin}(2\pi \xi_{2v}), \nonumber \\
&v_{1,e}^{\prime} = \sqrt{1 - v_{3,e}^{\prime 2}}\ {\rm cos}(2\pi \xi_{2v}). \nonumber
%\end{align}
\end{eqnarray}
The momentum of the electron is ${\bf p_{e}^{\prime}} = \gamma_e m_e \beta_e c\ (v_{1,e}^{\prime}\hat{x} + v_{2,e}^{\prime}\hat{y} + v_{3,e}^{\prime}\hat{z})$. The momentum of the outgoing positron, ${\bf p_{pos}^{\prime}} = \gamma_{pos} m_e \beta_{pos} c\ (v_{1,pos}^{\prime}\hat{x} + v_{2,pos}^{\prime}\hat{y} + v_{3,pos}^{\prime}\hat{z})$, is obtained using conservation of momentum in each direction with the expression
\begin{eqnarray}
v_{i,pos}^{\prime} = \frac{(E_{\gamma,1}^{\prime}/c)\Omega_{i,1}^{\prime} + (E_{\gamma,2}^{\prime}/c)\Omega_{i,2}^{\prime} - \gamma_e m_e \beta_e c v_{i,e}^{\prime}}{\gamma_{pos}m_e \beta_{pos} c}, \nonumber
\end{eqnarray} 
where the index $i=1,2,3$ denotes the component of the electron/photon direction vector.

 \subsection{Pair annihilation}
The positrons generated from the pair production events are stored in an array with their energies and directions. After a scattering event occurs, if the number of positrons in this array is non-zero, a positron with energy $\gamma_{pos}$ and direction $(v_{1,pos}^{\prime},v_{2,pos}^{\prime},v_{3,pos}^{\prime})$ is randomly selected for the annihilation process. An electron with energy $\gamma_{e}$ and direction $(v_{1,e}^{\prime},v_{2,e}^{\prime},v_{3,e}^{\prime})$ is drawn independently for scattering with photon based on its scattering probability. Next we evaluate the Lorentz factor of the positron in the comoving frame of the electron, $\gamma_{r} = (\gamma_{pos} - \gamma_e)/(1 - \gamma_{pos} \gamma_e /c^2)$. The pair annihilation cross section can be evaluated as (see \citealt{Poz83})
\begin{eqnarray}
\sigma_{a}(\gamma_{r}) = \frac{\pi r_{e}^{2}}{(\gamma_{r}+1)} \nonumber \\
 \left[\left(\frac{\gamma_{r}^2 + 4\gamma_{r} + 1}{\gamma_{r}^2 - 1}\right){\rm ln}(\gamma_{r} + \sqrt{\gamma_{r}^2 - 1}) - \frac{\gamma_{r}+3}{\sqrt{\gamma_{r}^2 -1}}\right].
 \end{eqnarray}
A random number $\zeta_{a}$ is then drawn and pair annihilation event occurs only if $\zeta_{a} \leq \sigma_{a}/\sigma_{T}$ is satisfied. 

Two outgoing photons are generated from the selected electron and positron after every pair annihilation event. In order to determine the energies of the outgoing photons, we draw a random number $\zeta_{aE}$ and assign their energies to be: $E_{\gamma,1}^{\prime} = \zeta_{aE} m_e c^2 (\gamma_{e} + \gamma_{pos})$ and $E_{\gamma,2}^{\prime} = (1 - \zeta_{aE}) m_e c^2 (\gamma_{e} + \gamma_{pos})$. The direction of the first photon $(\Omega_{1,1}^{\prime},\Omega_{2,1}^{\prime},\Omega_{3,1}^{\prime})$ is fixed with random numbers $\zeta_{1v}$ and $\zeta_{2v}$ 
\begin{eqnarray}
&\Omega_{3,1}^{\prime} = 2\zeta_{1v} - 1, \nonumber \\
&\Omega_{2,1}^{\prime} = \sqrt{1 - \Omega_{3,1}^{\prime 2}}\ {\rm sin}(2\pi \zeta_{2v}), \nonumber \\
&\Omega_{1,1}^{\prime} = \sqrt{1 - \Omega_{3,1}^{\prime 2}}\ {\rm cos}(2\pi \zeta_{2v}), \nonumber 
\end{eqnarray}
and its momentum is ${\bf p_{\gamma,1}^{\prime}} = (E_{\gamma,1}^{\prime}/c)(\Omega_{1,1}^{\prime} \hat{x} + \Omega_{2,1}^{\prime} \hat{y} + \Omega_{3,1}^{\prime} \hat{z})$. The momentum of the second photon ${\bf p_{\gamma,2}^{\prime}} = (E_{\gamma,2}^{\prime}/c)(\Omega_{1,2}^{\prime} \hat{x} + \Omega_{2,2}^{\prime} \hat{y} + \Omega_{3,2}^{\prime} \hat{z})$ is obtained from momentum conservation in each direction
\begin{eqnarray}
\Omega_{j,2}^{\prime} = \frac{(\gamma_{e}\beta_{e}m_e c^2 v_{j,e}^{\prime} + \gamma_{pos}\beta_{pos}m_e c^2 v_{j,pos}^{\prime} - E_{\gamma,1}^{\prime}\Omega_{j,1}^{\prime})}{E_{\gamma,2}^{\prime}},
\end{eqnarray}
where the index $j = 1,2,3$ denotes the component of the photon direction vector.

\section{Pair number density at equilibrium}
\label{Appendix_pair}
In this Appendix, we estimate the number of electron-positron pairs when the pair production and annihilation processes reach equilibrium and compare it with the total number of electrons present initially in the jet. We first calculate the number of photons with sufficient energy in the lab frame in order to generate pairs, $E_{\gamma,pair} \sim \Gamma m_e c^2 \sim 1.5\times10^{4}\ {\rm keV}$ (for $\Gamma=30$). The number of photons within a given energy range $(E_{a}^{\prime},E_{b}^{\prime})$ can be written in terms of the specific photon flux $f_{\nu}$ as, $N_{\gamma} = A \int_{E_{a}^{\prime}}^{E_{b}^{\prime}} (f_{\nu}/\nu)d\nu$. Here $A$ is a normalization constant that is determined from the total photon number $N_{\gamma,tot} = 2\times10^{7}$ as well as the shape of the photon spectrum. For a typical photon spectrum as shown in Figure \ref{fig7}, the minimum, peak, pair production and maximum photon energies in the jet-comoving frame are $E_{\gamma,min}^{\prime} \sim 3.33\times10^{-3}\ {\rm keV}$, $E_{\gamma,peak}^{\prime} \sim 3.33\times10^{1}\ {\rm keV}$, $E_{\gamma,pair}^{\prime} \sim 5.00\times10^{2}\ {\rm keV}$ and $E_{\gamma,max}^{\prime} \sim 1.67\times10^{4}\ {\rm keV}$, respectively. As $f_{\nu} \propto \nu^0$ for $E_{\gamma} < E_{\gamma,peak}$ and $f_{\nu} \propto \nu^{-1.35}$ for $E_{\gamma} > E_{\gamma,peak}$ from Figure \ref{fig7}, the normalization factor $A = 2.17\times10^{6}$ for $N_{\gamma,tot} = 2\times10^{7}$. The number of photons with sufficient energy required to produce pairs is then found to be, $N_{\gamma,pair} = (2.17\times10^{6})\int_{E_{\gamma,pair}^{\prime}}^{E_{\gamma,max}^{\prime}} (\nu^{-1.35}/\nu)d\nu \approx 3.62\times10^{2}$. 

The pair production optical depth is, $\tau_{\gamma\gamma} \sim (N_{\gamma,pair}\sigma_{\gamma\gamma,avg})/(4\pi R^2)$, where the average pair production cross section $\sigma_{\gamma\gamma,avg} = \int_{y_{min}}^{y_{max}}\sigma_{\gamma\gamma}(y)(f_{y}/y)dy\big{/}$ $\int_{y_{min}}^{y_{max}}(f_{y}/y)dy$ for $\sigma_{\gamma\gamma}$ given by Equation \ref{pp_cs}. While $y_{min}=1$ and $y_{max} \sim (1/\sqrt{2})(E_{\gamma,max}^{\prime}/m_e c^2) \sim 25$ for photon spectrum with $f_{y} \sim y^{-1.35}$, $\langle\rm{cos\ }\theta=0\rangle$ for photons with isotropic distribution in the jet-comoving frame. Substituting these and performing the integral gives $\sigma_{\gamma\gamma,avg} \approx 0.168 \sigma_{T}$ and pair production optical depth is
\begin{eqnarray}
\tau_{\gamma\gamma} = \frac{N_{\gamma,pair}\sigma_{\gamma\gamma,avg}}{4\pi R^2} = \left(\frac{N_{\gamma,pair}}{N_{e,tot}}\right)\times 0.168\tau_{e} \approx 0.304\tau_{e},
\end{eqnarray}
where $N_{e,tot}$ is the total electron number in the jet and $\tau_{e} = (N_{e,tot}\sigma_{T})/(4\pi R^2)$ is the electron scattering optical depth. Similarly, the pair annihilation optical depth can be written as 
\begin{eqnarray}
\tau_{e^{-}e^{+}} = \frac{N_{e^{-}e^{+}}\sigma_{a}}{4\pi R^2} \approx \frac{3}{8}\frac{N_{e^{-}e^{+}}}{N_{e,tot}}\frac{\tau_{e}}{\beta_{e}},
\end{eqnarray} 
where $N_{e^{-}e^{+}}$ is the number of pairs in the jet and $\sigma_{a} \approx (3/8)(\sigma_{T}/\beta_{e})$ is the asymptotic pair annihilation cross section for sub-relativistic electrons. As the pair production and annihilation rates match at equilibrium $\tau_{\gamma\gamma}c = \tau_{e^{-}e^{+}}\beta_{e}c$, which further gives
\begin{eqnarray}
N_{e^{-}e^{+}} = \frac{8}{3}\frac{\sigma_{\gamma\gamma,avg}}{\sigma_{T}}N_{\gamma,pair} \approx 0.8 N_{e,tot}.\nonumber
\end{eqnarray}
Therefore, the number of electron-positron pairs in the jet is approximately equal to the electrons initially present in the jet. As the Comptonized output photon spectrum can get affected by the increase in electron number density in the jet, particularly at large values of optical depth, it is important to consider pair processes for our MCRaT simulations.

\end{appendix}

\label{lastpage}

\end{document}